\documentclass[aps,prd,reprint,superscriptaddress,nofootinbib,longbibliography]{revtex4-2}

\usepackage{amsmath,amssymb,bm}
\usepackage{booktabs}
\usepackage{graphicx}
\usepackage{xcolor}
\usepackage{pgfplots}
\usepackage{hyperref}
\usepackage{tikz}
\pgfplotsset{compat=1.18}
\usepgfplotslibrary{groupplots}
\usetikzlibrary{arrows.meta,positioning,calc,decorations.pathmorphing,decorations.markings}

\newcommand{\dd}{\mathrm{d}}
\newcommand{\Tr}{\operatorname{Tr}}
\newcommand{\MSbar}{\overline{\mathrm{MS}}}
\newcommand{\BZ}{\mathrm{BZ}}
\newcommand{\SUtwo}{\mathrm{SU}(2)}

\newcommand{\DeltaGamma}{\Delta\Gamma}
\newcommand{\arcE}[3]{\draw[#1] (#2:1) arc (#2:#3:1);}
\newcommand{\dotv}[1]{\fill (#1:1) circle (2pt);}
\newcommand{\ctv}[1]{\draw[fill=white] (#1) circle (3.5pt); \draw ($(#1)+(-2.5pt,-2.5pt)$) -- ($(#1)+(2.5pt,2.5pt)$); \draw ($(#1)+(-2.5pt,2.5pt)$) -- ($(#1)+(2.5pt,-2.5pt)$);}
\newcommand{\legn}[2]{\draw[higgs] (#1:1) -- ++(#2:0.95);}
\newcommand{\pbox}[1]{\node[draw,rounded corners,align=center,font=\scriptsize,inner sep=3pt,#1]}

\tikzset{
  higgs/.style={thick},
  gold/.style={thick,dashed},
  gauge/.style={thick,decorate,decoration={snake,amplitude=1.1pt,segment length=5pt}},
  scal/.style={thick,postaction={decorate},decoration={markings,mark=at position 0.55 with {\arrow{Stealth}}}},
  ghost/.style={thick,densely dotted,postaction={decorate},decoration={markings,mark=at position 0.55 with {\arrow{Stealth}}}},
  scalarline/.style={draw=red!70!black,thick},
  goldline/.style={draw=orange!80!black,thick},
  gaugelink/.style={draw=blue!65!black,thick},
  ghostline/.style={draw=violet!70!black,thick,densely dashed},
  completebox/.style={draw=green!50!black,fill=green!8,rounded corners=2pt,
    align=center,minimum height=8mm,text width=3.0cm,font=\small},
  pilotbox/.style={draw=orange!70!black,fill=orange!9,rounded corners=2pt,
    align=center,minimum height=8mm,text width=3.0cm,font=\small},
  openbox/.style={draw=gray!70!black,fill=gray!10,rounded corners=2pt,
    align=center,minimum height=8mm,text width=3.0cm,font=\small}
}

\begin{document}

\preprint{Companion manuscript, November 19, 2026}
\title{Higgs Fluctuation Solver in Four Dimensions: Part II---Toward Finite
Lattice-to-$\MSbar$ Matching in the $\SUtwo$--Higgs Extension}

\author{M. S. Cook}
\email{merrittscook1955@gmail.com}
\homepage[]{https://merrittscook1955-ctrl.github.io/in-vero-vacuo/}
\affiliation{In Vero Vacuo LLC}
\date{November 19, 2026}

\begin{abstract}
We present a lattice-to-$\MSbar$ matching framework for the $SU(2)$--Higgs
theory, including Wilson-plaquette and fundamental scalar-link actions, a
counterterm basis through $\rho^{12}$, audited momentum-dependent lattice
vertices, and fixed routings and projectors. At the benchmark $a=0.8$,
$m_0^2=\mu_R^2=0.4$, $g_0=0.65$, and $(\lambda,\gamma,\delta)=(-2,0.8,0.08)$,
the potential is $V_0(\rho)=\frac15\rho^2-\frac12\rho^4+
\frac{2}{15}\rho^6+\frac1{100}\rho^8$. The scalar-only radial matching
shifts, ordered as $(\delta m^2,\delta\lambda,\delta\gamma,\delta\delta,
\delta c_{10},\delta c_{12})$, are
$(2.75293,-1.45119,-2.85120,0.356724,1.07432,6.20492)$ in solver units,
with the continuum-renormalized-minus-lattice convention. The Higgs two-point slope gives
$\delta Z_H=0.0329117974\pm0.00000036$ (QMC error). We also evaluate
gauge-linked four-, six-, and eight-Higgs projections and derive the continuum
one-loop gauge two-point tensor. The tensor passes 29 internal checks; at
$p^2=0.4$, its total transverse projection is $\Pi_T=-0.01847$ at zero
regulator and $-0.01716$ at $\lambda_{\rm IR}/\mu_R=1/8$, while the
regulator-induced longitudinal projection is $\Pi_L=1.04\times10^{-5}$ at
$\lambda_{\rm IR}/\mu_R=1/32$. These results establish a tested foundation
and quantitative benchmarks for the matching program. The gauge-linked
projections remain preliminary, and full gauge-sector matching and
Ward--Slavnov--Taylor closure are needed before a matched $SU(2)$ decay
prefactor can be claimed.

\end{abstract}

\maketitle

\section{Introduction}
\label{sec:introduction}

Semiclassical false-vacuum decay depends on both a Euclidean saddle and the
renormalized fluctuation determinant around that saddle
\cite{Coleman1977,CallanColeman1977}. Part I developed a numerical pipeline
for a four-component global-$O(4)$ scalar theory: continuum bounce shooting,
relaxation of the sampled profile on a periodic four-dimensional lattice,
matrix-free Hessian operations, collective-mode analysis, and complementary
bare-lattice and continuum-subtracted determinant calculations
\cite{CookPartI}. The scalar calculation is a controlled benchmark, not a
Standard Model prediction. In particular, its retained scalar potential is
not closed under the one-loop determinant without higher local operators,
and its finite $\rho^{10}$ and $\rho^{12}$ coefficients were not matched.

The metastability of the Standard Model Higgs vacuum has been studied using
renormalization-group evolution and semiclassical decay calculations,
including gauge-invariant one-loop treatments
\cite{Degrassi2012,Buttazzo2013,Chigusa2018}. Numerical continuum methods
have also computed fluctuation prefactors for scalar theories in four
spacetime dimensions \cite{DunneMin2005}; consequently, the motivation here
is not that four-dimensional prefactors are wholly unknown. Euclidean lattice
Monte Carlo methods for vacuum-decay rates have, for example, been tested in
quantum-mechanical systems \cite{Shen2023}. The narrower objective of this
companion is to develop and test components of finite-lattice-to-$\MSbar$
matching for the $\SUtwo$--Higgs theory, as an enabling step toward a
gauge-coupled fluctuation determinant. It does not claim a completed
nonperturbative electroweak decay-rate calculation. Recent analytic studies
have refined false-vacuum prefactors and gauge-sector treatments, including
determinant methods and gradient corrections
\cite{GuadaNemevsek2020,AiCruzGarbrechtTamarit2020,BaratellaGauge2025}.

The present paper follows the next computational question: how should the
finite lattice regulator be related to continuum $\MSbar$ parameters when
the scalar is coupled to a non-Abelian gauge field? This is not accomplished
by replacing continuum momenta with lattice momenta in a final determinant.
Each coefficient in the local action must be fixed by comparing
renormalized continuum and lattice amplitudes at identical external
kinematics. The sign convention used below is
\begin{equation}
  \DeltaGamma^{(n)}(p_i)\equiv
  \Gamma^{(n)}_{\MSbar,\,\mathrm{ren}}(p_i)
  -\Gamma^{(n)}_{\mathrm{lat}}(p_i).
  \label{eq:matching-difference}
\end{equation}
Here, $n$ is the number of external legs, so $\Gamma^{(n)}$ denotes an
$n$-point one-particle-irreducible vertex, and $p_i$ are the incoming
Euclidean external momenta specifying the matching kinematics. The
subscripts $\MSbar,\,\mathrm{ren}$ denote the continuum vertex renormalized
in the $\MSbar$ scheme, while $\mathrm{lat}$ denotes the corresponding
lattice-regularized vertex; both are evaluated at the same specified
kinematics and in the same field convention. Thus $\DeltaGamma^{(n)}$ is a
signed continuum-minus-lattice difference: a positive value means the
continuum-renormalized vertex exceeds its lattice counterpart at those
kinematics. Its local projections determine the corresponding counterterms.
The difference contains a scheme- and regulator-dependent finite part; it
is not a universal constant independent of the chosen lattice action,
spacing, subtraction scheme, and matching momenta
\cite{Wilson1974,LW,Reisz,Kawai}.

The gauge extension also introduces consistency checks absent from a
scalar-only calculation. Scalar--gauge mixing, seagull vertices, pure-gauge
and ghost loops, link-coordinate measure terms, and the Ward--Slavnov--Taylor
identities must be treated in a common convention. A collection of local
amplitudes that individually looks stable is not a valid matched action if
the gauge two-point function and its identities do not close
\cite{Abbott,Taylor}. Likewise, a scalar one-loop determinant cannot be
expanded into a finite set of polynomial counterterms indefinitely: in the
retained truncation, $\rho^{10}$ and $\rho^{12}$ terms are already required.

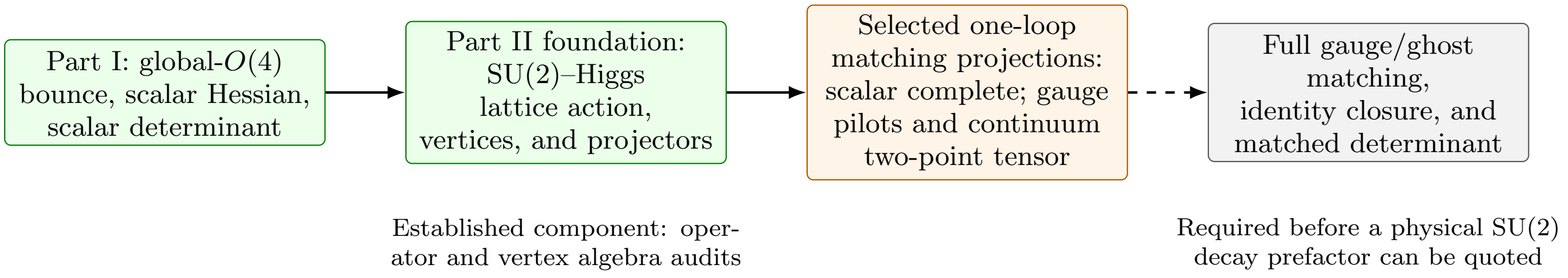
\begin{figure*}[t]
\centering
\begin{tikzpicture}[>=Latex,node distance=8mm and 9mm]
  \node[completebox,text width=3.4cm] (partone) {Part I: global-$O(4)$\\bounce, scalar Hessian,\\scalar determinant};
  \node[completebox,right=of partone,text width=3.4cm] (action) {Part II foundation:\\$\SUtwo$--Higgs lattice action,\\vertices, and projectors};
  \node[pilotbox,right=of action,text width=3.4cm] (loops) {Selected one-loop\\matching projections:\\scalar complete; gauge pilots and continuum two-point tensor};

  \node[openbox,right=of loops,text width=3.4cm] (closure) {Full gauge/ghost matching,\\identity closure, and\\matched determinant};
  \draw[-{Latex[length=2mm]},thick] (partone) -- (action);
  \draw[-{Latex[length=2mm]},thick] (action) -- (loops);
  \draw[-{Latex[length=2mm]},thick,dashed] (loops) -- (closure);
  \node[below=5mm of action,align=center,font=\footnotesize,text width=5.5cm] {
    Established component: operator and vertex algebra audits};
  \node[below=5mm of closure,align=center,font=\footnotesize,text width=5cm] {
    Required before a physical $\SUtwo$ decay prefactor can be quoted};
\end{tikzpicture}
\caption{Program map. Part I supplies the scalar benchmark, while this paper
records the lattice-to-continuum matching infrastructure and its partial
one-loop tests. The dashed transition marks work that is not complete; the
pilot amplitudes below are not matching coefficients for a production
calculation.}
\label{fig:program-map}
\end{figure*}

This manuscript is deliberately a methods-and-status companion rather than
a claim of completed full-$\SUtwo$ matching. Its concrete contributions are:
\begin{enumerate}
\item a fixed lattice action and field convention, with the local operator
  basis required by the scalar one-loop sector;
\item exact momentum-dependent vertex generators and a record of their
  algebraic validation;
\item fixed kinematic routings and projectors for two-point and local
  $2n$-point matching;
\item results for the scalar-only radial coefficients, the Higgs kinetic
  coefficient, and selected gauge-linked pilot projections; and
\item a separation of numerical integration errors from regulator-model
  uncertainty and still-omitted physics.
\end{enumerate}
The first two items establish the computational foundation. The latter
results do not populate the full coefficient set: although the continuum
gauge two-point tensor is now available (Sec.~\ref{sec:g2pt}), its independent
audit, lattice comparison, and associated identity checks remain open.
Figure~\ref{fig:program-map} summarizes the relationship between Parts I and
II and the remaining path to a physical rate.

The paper is organized as follows. Section~\ref{sec:model} defines the
continuum and lattice theories. Section~\ref{sec:vertices} describes the
vertex generators and their audits. Section~\ref{sec:matching-method}
specifies the kinematics, infrared regulator, and projectors.
Section~\ref{sec:results} presents the numerical results and their
limitations. Section~\ref{sec:g2pt} derives and validates the continuum
gauge two-point tensor. Sections~\ref{sec:discussion} and
~\ref{sec:conclusion} summarize the current status and the gates required
for completion. Appendices A and B define the symbols, variables, diagram
connections, and free propagators. Appendix C contains a project status
dashboard.

\section{Model, lattice action, and operator basis}
\label{sec:model}

\subsection{Fields and scalar potential}
\label{subsec:fields}

We write the complex fundamental doublet in terms of four real fields as
\begin{equation}
 H_x=\frac{1}{\sqrt{2}}
 \begin{pmatrix}
  \phi_x^2+i\phi_x^1\\
  \phi_x^4-i\phi_x^3
 \end{pmatrix},\qquad
 \rho_x^2=2H_x^\dagger H_x=\sum_{A=1}^{4}(\phi_x^A)^2.
 \label{eq:doublet}
\end{equation}
At vanishing gauge field, this convention gives canonical kinetic terms for
the four real components and makes the global-$O(4)$ limit explicit. Around
a radial background, $\phi^4=\bar\rho+h$ and $\phi^a=\chi^a$ ($a=1,2,3$),
where $h$ is the radial fluctuation and $\chi^a$ are the Goldstone
fluctuations.

The tree potential and the counterterm potential are taken to be
\begin{align}
 V_0(\rho)&=-\frac{1}{2}\mu_{\rm sq}\rho^2+
  \frac{\lambda}{4}\rho^4+\frac{\gamma}{6}\rho^6+
  \frac{\delta}{8}\rho^8,\label{eq:potential-tree}\\
 V_{\rm ct}(\rho)&=\frac{1}{2}\delta m^2\rho^2+
  \frac{1}{4}\delta\lambda\rho^4+
  \frac{1}{6}\delta\gamma\rho^6+
  \frac{1}{8}\delta\delta\rho^8\nonumber\\
 &\quad +\frac{1}{10}\delta c_{10}\rho^{10}+
  \frac{1}{12}\delta c_{12}\rho^{12}.
 \label{eq:potential-ct}
\end{align}
The false-vacuum mass parameter is $m_0^2=-\mu_{\rm sq}$. Since the real
scalar has canonical dimension one in four dimensions, the coefficients of
$\rho^{10}$ and $\rho^{12}$ have mass dimensions $-6$ and $-8$,
respectively. They are included as perturbative matching counterterms for
the specified truncation; they are not asserted to be additional leading
tree interactions. This distinction is important when interpreting the
finite coefficients in Table~\ref{tab:scalar-matching}.

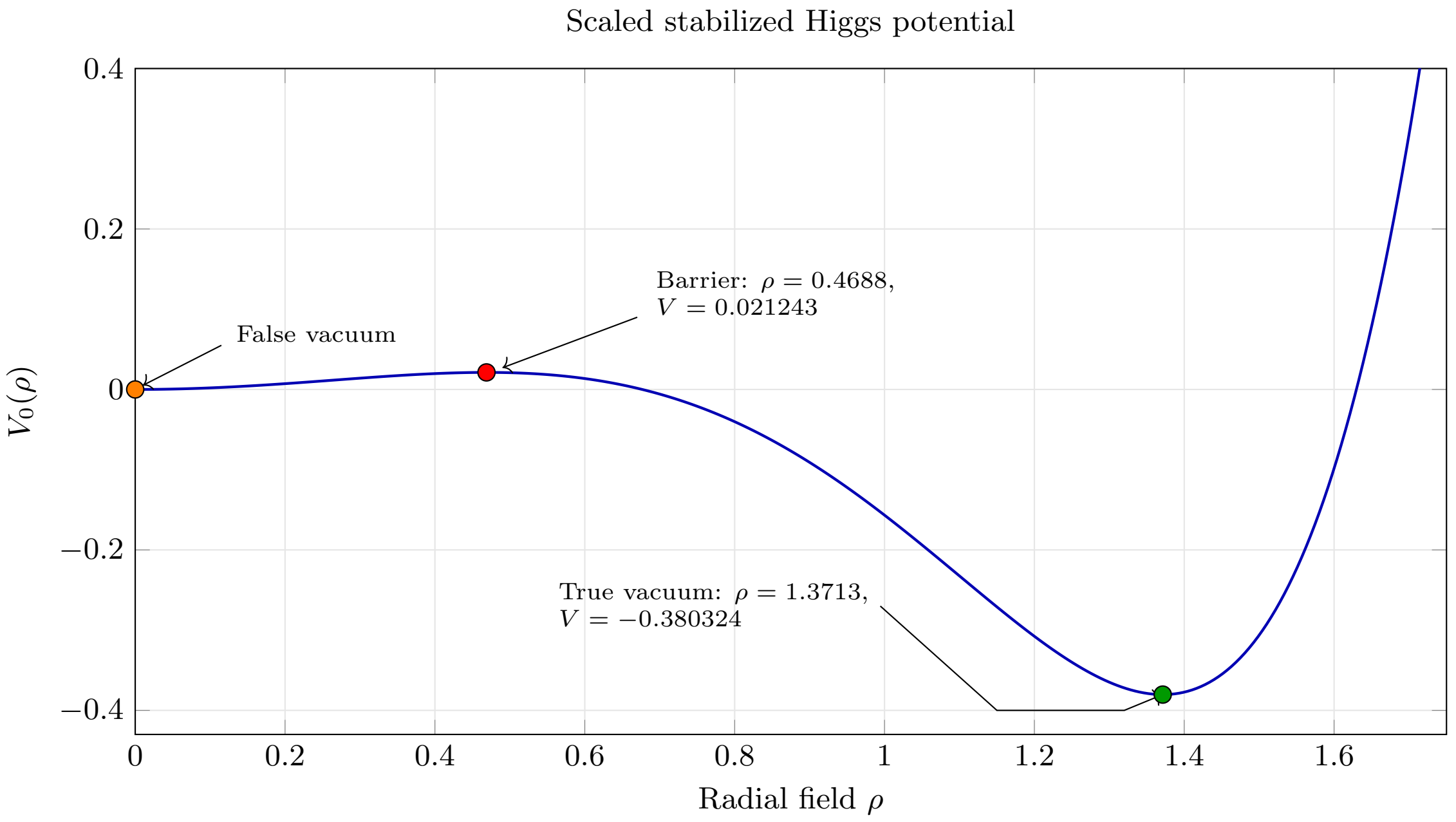
\begin{figure*}[t]
\centering
\begin{tikzpicture}
\begin{axis}[
  width=0.84\textwidth,
  height=0.47\textwidth,
  title={Scaled stabilized Higgs potential},
  xlabel={Radial field $\rho$},
  ylabel={$V_0(\rho)$},
  xmin=0,xmax=1.75,
  ymin=-0.43,ymax=0.4,
  grid=major,
  grid style={gray!20},
  tick label style={font=\small},
  label style={font=\small},
  clip=true
]
\addplot[blue!70!black,thick,domain=0:1.75,samples=300]
  {0.2*x^2-0.5*x^4+(2/15)*x^6+(1/100)*x^8};
\addplot[only marks,mark=*,mark size=2.5pt,
  mark options={fill=orange,draw=black}] coordinates {(0,0)};
\addplot[only marks,mark=*,mark size=2.5pt,
  mark options={fill=red,draw=black}] coordinates {(0.4688,0.021243)};
\addplot[only marks,mark=*,mark size=2.5pt,
  mark options={fill=green!60!black,draw=black}]
  coordinates {(1.3713,-0.380324)};
\node[anchor=west,font=\scriptsize] at (axis cs:0.12,0.07)
  {False vacuum};
\draw[->] (axis cs:0.115,0.055) -- (axis cs:0.01,0.005);
\node[anchor=west,align=left,font=\scriptsize] at (axis cs:0.68,0.12)
  {Barrier: $\rho=0.4688$,\\$V=0.021243$};
\draw[->] (axis cs:0.67,0.09) -- (axis cs:0.49,0.027);
\node[anchor=west,align=left,font=\scriptsize] (truevaclabel)
  at (axis cs:0.55,-0.27)
  {True vacuum: $\rho=1.3713$,\\$V=-0.380324$};
\draw[->] (truevaclabel.east) -- (axis cs:1.15,-0.40)
  -- (axis cs:1.32,-0.40) -- (axis cs:1.3713,-0.380324);
\end{axis}
\end{tikzpicture}
\caption{Scaled benchmark Higgs potential at $L=20$ with
$m_0^2=0.4$, $\lambda=-2$, $\gamma=0.8$, and $\delta=0.08$. The false
vacuum is at $\rho=0$, the barrier at $(\rho,V)\simeq(0.4688,0.021243)$,
and the true minimum at $(\rho,V)\simeq(1.3713,-0.380324)$. A common
rescaling of all potential coefficients preserves these field locations
while rescaling the potential energies. Benchmark parameters and extrema
are from Ref.~\cite{CookPartI}.}
\label{fig:benchmark-potential}
\end{figure*}

For the plotted benchmark, $m_0^2=-\mu_{\rm sq}=0.4$,
$\lambda=-2$, $\gamma=0.8$, and $\delta=0.08$. The origin is a local false
vacuum; a barrier at $\rho\simeq0.4688$ separates it from the lower true
minimum at $\rho\simeq1.3713$. Figure~\ref{fig:benchmark-potential} shows
the stabilized tree-level shape. The matching counterterms are not included
in this plot; they are finite scalar-only differences at the stated lattice
spacing and subtraction scale.

The scalar operators required by the calculation are summarized in
Table~\ref{tab:operator-basis}. The wave-function coefficient is kept
separately from the local potential coefficients.

\begin{table}[t]
\centering
\small
\begin{tabular}{@{}lll@{}}
\toprule
Operator & Coefficient & Mass dimension\\
\midrule
$(\partial_\mu\rho)^2/2$ & $\delta Z_H$ & $0$\\
$\rho^2/2$ & $\delta m^2$ & $2$\\
$\rho^4/4$ & $\delta\lambda$ & $0$\\
$\rho^6/6$ & $\delta\gamma$ & $-2$\\
$\rho^8/8$ & $\delta\delta$ & $-4$\\
$\rho^{10}/10$ & $\delta c_{10}$ & $-6$\\
$\rho^{12}/12$ & $\delta c_{12}$ & $-8$\\
\bottomrule
\end{tabular}
\caption{Local operator basis through $\rho^{12}$. The higher powers
complete the retained scalar one-loop counterterm basis; they do not imply
that a finite polynomial truncation is closed at arbitrary loop order.}
\label{tab:operator-basis}
\end{table}

\subsection{Gauge and scalar-link actions}
\label{subsec:lattice-action}

Let $T^a=\sigma^a/2$, and parametrize each link by
\begin{equation}
 U_{x\mu}=\exp[-iag_0A^a_{x\mu}T^a],
 \label{eq:link}
\end{equation}
with $A^a_{x\mu}$ associated with the link midpoint. The oriented plaquette
is
\begin{equation}
 U_{x,\mu\nu}=U_{x\mu}U_{x+a\hat\mu,\nu}
 U^\dagger_{x+a\hat\nu,\mu}U^\dagger_{x\nu},
\end{equation}
and the gauge action is the Wilson action
\begin{equation}
 S_g=\frac{4}{g_0^2}\sum_x\sum_{\mu<\nu}
 \left(1-\frac{1}{2}\operatorname{ReTr}U_{x,\mu\nu}\right).
 \label{eq:wilson-action}
\end{equation}
Its weak-field continuum normalization is
$\frac{1}{4}\int\dd^4x\,F^a_{\mu\nu}F^a_{\mu\nu}$
\cite{Wilson1974,LW}.

The gauge-covariant scalar kinetic term is the nearest-neighbor link action
\begin{equation}
 S_{H,\rm link}=a^4\sum_{x,\mu}
 \frac{\bigl(U_{x\mu}H_{x+a\hat\mu}-H_x\bigr)^\dagger
       \bigl(U_{x\mu}H_{x+a\hat\mu}-H_x\bigr)}{a^2}.
 \label{eq:scalar-link}
\end{equation}
At $U=1$, Eq.~\eqref{eq:scalar-link} reduces to
$\frac{a^4}{2}\sum_{x,\mu,A}[(\phi^A_{x+a\hat\mu}-\phi_x^A)/a]^2$.
The full perturbative action used to generate vertices is
\begin{equation}
\begin{aligned}
 S={}&S_g+S_{H,\rm link}\\
 &+a^4\sum_x[V_0(\rho_x)+V_{\rm ct}(\rho_x)]\\
 &+S_{\rm gf}+S_{\rm gh}+\delta Z_H S_{H,\rm link}.
\end{aligned}
\label{eq:full-action}
\end{equation}
The final term denotes the kinetic counterterm insertion, not an additional
independent tree action.

For a radial background with vanishing background gauge field, the
background $R_\xi$ gauge is defined by
\begin{align}
 F_x^a&=(DA)^a_x-\xi q\bar\rho_x\chi_x^a,
 &S_{\rm gf}&=\frac{a^4}{2\xi}\sum_{x,a}(F_x^a)^2,
 \label{eq:gauge-fixing}
\end{align}
where $q=g_0/2$ and $D$ is the lattice backward divergence. The ghost
operator is fixed by the variation of this same gauge-fixing function,
\begin{equation}
 M_{xy}^{ab}=\frac{\delta F_x^a}{\delta\omega_y^b},\qquad
 S_{\rm gh}=a^4\sum_{x,y}\bar c_x^aM_{xy}^{ab}c_y^b,
 \label{eq:ghost-action}
\end{equation}
under the infinitesimal gauge transformation of both $H$ and $U$. Thus the
ghost vertices are not chosen independently of the link and gauge-fixing
conventions.

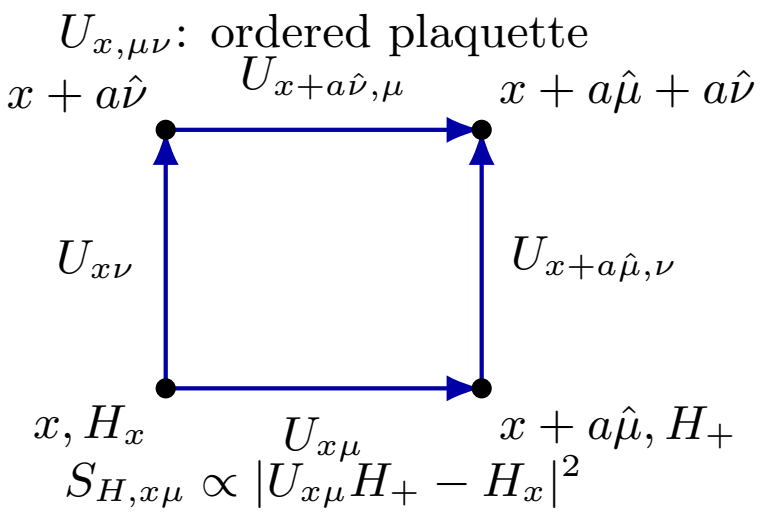
\begin{figure}[t]
\centering
\begin{tikzpicture}[scale=0.95,>=Latex]
  \coordinate (x) at (0,0);
  \coordinate (m) at (2.2,0);
  \coordinate (mn) at (2.2,1.8);
  \coordinate (n) at (0,1.8);
  \draw[gaugelink,->] (x) -- (m) node[midway,below=2pt] {$U_{x\mu}$};
  \draw[gaugelink,->] (m) -- (mn) node[midway,right=2pt] {$U_{x+a\hat\mu,\nu}$};
  \draw[gaugelink,->] (n) -- (mn) node[midway,above=2pt] {$U_{x+a\hat\nu,\mu}$};
  \draw[gaugelink,->] (x) -- (n) node[midway,left=2pt] {$U_{x\nu}$};
  \foreach \p in {x,m,mn,n}{\fill (\p) circle (2pt);}
  \node[below left] at (x) {$x,H_x$};
  \node[below right] at (m) {$x+a\hat\mu,H_+$};
  \node[above right] at (mn) {$x+a\hat\mu+a\hat\nu$};
  \node[above left] at (n) {$x+a\hat\nu$};
  \node at (1.1,-0.65) {$S_{H,x\mu}\propto|U_{x\mu}H_+-H_x|^2$};
  \node at (1.1,2.45) {$U_{x,\mu\nu}$: ordered plaquette};
\end{tikzpicture}
\caption{Link and plaquette conventions. Reversing an edge replaces its
matrix by the Hermitian conjugate. The scalar kinetic action uses each
oriented link through Eq.~\eqref{eq:scalar-link}; the ordered plaquette
generates the Wilson gauge vertices.}
\label{fig:lattice-links}
\end{figure}

Figure~\ref{fig:lattice-links} depicts the elementary lattice objects.
The link exponentials are also the source of the multi-gauge scalar
vertices: differentiating Eq.~\eqref{eq:scalar-link} $r$ times with
respect to $A$ at $A=0$ produces the scalar--scalar--$r$-gauge rule. The
corresponding derivatives of Eq.~\eqref{eq:wilson-action} generate pure
gauge vertices.

\subsection{Benchmark and free kernels}
\label{subsec:benchmark}

The benchmark parameters used throughout this paper are listed in
Table~\ref{tab:benchmark}. They define a convenient numerical test point,
not a determination of the bare-to-renormalized gauge coupling
or a physical electroweak parameter set. In particular, $g_0=0.65$ is held
fixed in the lattice link variables; the spacing dependence of $g_0(a)$ has
not been matched.

\begin{table}[t]
\centering
\small
\begin{tabular}{@{}lll@{}}
\toprule
Quantity & Definition & Benchmark\\
\midrule
$a$ & lattice spacing & $0.8$\\
$m_0^2$ & $-\mu_{\rm sq}$ & $0.4$\\
$\mu_R^2$ & $\MSbar$ scale squared & $0.4$\\
$\lambda,\gamma,\delta$ & tree potential & $-2,\ 0.8,\ 0.08$\\
$g_0$ & bare lattice gauge coupling & $0.65$\\
$q$ & $g_0/2$ & $0.325$\\
$\xi$ & background gauge parameter & $1$\\
$\bar\rho$ & local-vertex diagnostic background & $0.37$\\
\bottomrule
\end{tabular}
\caption{Benchmark convention. The subtraction scale is
$\mu_R=\sqrt{0.4}=0.6324555\ldots$; it is distinct from external momentum
$p$.}
\label{tab:benchmark}
\end{table}

For $\hat k_\mu=(2/a)\sin(ak_\mu/2)$, the scalar free propagator at the
false vacuum is $(\hat k^2+m_0^2)^{-1}$. In background Feynman gauge and
with the common auxiliary infrared shift introduced below, the massless
gauge and ghost denominators are $\hat k^2+\lambda_{\rm IR}^2$; the
continuum denominators use $k^2+\lambda_{\rm IR}^2$. The same shift is used
on both sides of Eq.~\eqref{eq:matching-difference} before extrapolation.
The scalar masses are not shifted.

\section{Vertex generators and algebraic validation}
\label{sec:vertices}

\subsection{Expansion around a radial background}
\label{subsec:quadratic-check}

For $H=(\chi,\bar\rho+h)^T/\sqrt{2}$ in a single real generator plane,
expansion of Eq.~\eqref{eq:scalar-link} to quadratic order gives
\begin{align}
 S_{H,\rm link}^{(2)}&=\frac{a^4}{2}\sum_{x,\mu}
 \left[\frac{(\chi_{x+a\hat\mu}-\chi_x)^2}{a^2}
 +q^2\bar\rho_x\bar\rho_{x+a\hat\mu}A_{x\mu}^2\right]\nonumber\\
 &\quad+a^4\sum_{x,\mu}\frac{q}{a}A_{x\mu}
   (\bar\rho_{x+a\hat\mu}\chi_x-\bar\rho_x\chi_{x+a\hat\mu})
 +\text{radial terms}.
 \label{eq:quadratic-link}
\end{align}
This identifies the Goldstone kinetic term, the link seagull, and the
derivative $hA\chi$ mixing vertex. At a constant background, the Goldstone
and gauge quadratic terms combine with gauge fixing as required by the
chosen $R_\xi$ convention. At $\bar\rho=0$, the Higgs and three Goldstones
have mass $m_0$, while the gauge field and ghosts are massless before the
auxiliary infrared regulator is applied.

The exact nonlinear generators are obtained by differentiating the
exponential link and plaquette actions, rather than by replacing their
momentum dependence with a continuum approximation. For example,
\begin{equation}
 \left.\frac{\partial^r}{\partial A_{\mu_1}^{a_1}\cdots
 \partial A_{\mu_r}^{a_r}}
 \left(U_{x\mu}H_+-H_x\right)^\dagger
 \left(U_{x\mu}H_+-H_x\right)\right|_{A=0}
 \label{eq:link-generator}
\end{equation}
generates all scalar-link vertices with two scalar legs and $r$ gauge legs.
The color matrices remain ordered; this ordering matters for non-Abelian
vertices and is retained in the stored tensors. The same principle applies
to the Wilson action, whose derivatives yield the three- and four-gauge
vertices and their lattice momentum phases.

The potential vertices are derivatives of the invariant $V_0(\rho)$ and
$V_{\rm ct}(\rho)$ with respect to the real components $\phi^A$. They are
therefore fully symmetric in their scalar indices and are evaluated at the
requested background. For constant $\rho$, the radial and transverse
curvatures entering the scalar Hessian are
\begin{align}
 M_L^2(\rho)&=V_0''(\rho)
 =m_0^2+3\lambda\rho^2+5\gamma\rho^4+7\delta\rho^6,
 \label{eq:radial-mass}\\
 M_T^2(\rho)&=\frac{V_0'(\rho)}{\rho}
 =m_0^2+\lambda\rho^2+\gamma\rho^4+\delta\rho^6.
 \label{eq:goldstone-mass}
\end{align}
For the gauge boson mass parameter we use $M_A^2=q^2\rho^2$.

\subsection{Gauge fixing, ghosts, and link-coordinate measure}
\label{subsec:ghosts}

The Faddeev--Popov vertices follow from the exact variation in
Eq.~\eqref{eq:ghost-action}. This gives the expected free ghost kernel in
the symmetric background and the background-dependent ghost mass term in
the $R_\xi$ gauge. The generated tensors are checked against finite gauge
transformations of both the scalar and the links. Such checks test the
implemented gauge transformation and its derivatives; they do not establish
that a truncated numerical one-loop set obeys the continuum Ward identities.

The logarithmic link coordinates in Eq.~\eqref{eq:link} also carry a Haar
measure. For $\SUtwo$, the local density is
\begin{equation}
\begin{aligned}
 J(A)&=\left[\frac{\sin(ag_0|A|/2)}{ag_0|A|/2}\right]^2,\\
 -\log J(A)&=\frac{(ag_0)^2}{12}A^aA^a+O(A^4).
\end{aligned}
 \label{eq:haar}
\end{equation}
At quadratic order this contributes
$\Delta\Pi^{\rm Haar}_{\mu\nu}=g_0^2\delta_{\mu\nu}/(6a^2)$, equal to
$0.110026$ at the benchmark. It is a local measure term, not a propagating
loop. It must be included consistently in a gauge two-point matching
calculation; its inclusion alone does not repair an incomplete gauge
identity test.

\subsection{Local validation of the generated rules}
\label{subsec:vertex-validation}

The vertex module was checked at the benchmark against Taylor expansions,
finite transformations, momentum conservation, index permutations, and the
quadratic lattice kernels. Table~\ref{tab:vertex-audit} collects
representative maximum absolute discrepancies reported in these checks.
They are algebraic implementation diagnostics and are not statistical
uncertainties on any loop amplitude.

\begin{table*}[t]
\centering
\small
\begin{tabular}{@{}p{0.46\textwidth}p{0.19\textwidth}p{0.25\textwidth}@{}}
\toprule
Audit & Maximum discrepancy & What it tests\\
\midrule
Momentum conservation across generated tables & $0$ & Vertex routing\\
Scalar-link Taylor coefficients, orders 1--2 & $8.020\times10^{-9}$ & Link exponential derivatives\\
Complete scalar-link momentum/color table & $6.204\times10^{-9}$ & All recorded tensor components\\
Wilson-plaquette Taylor coefficients, orders 2--4 & $1.719\times10^{-9}$ & Pure-gauge expansion\\
Wilson leg-permutation symmetry & $4.442\times10^{-16}$ & Bose symmetry\\
Wilson quadratic kernel vs. transverse Hessian & $7.216\times10^{-16}$ & Free gauge normalization\\
Log-link gauge variation & $5.954\times10^{-11}$ & Infinitesimal link transformation\\
Scalar-link and plaquette gauge-invariance checks & $1.776\times10^{-15}$ / $2.220\times10^{-16}$ & Gauge transformation algebra\\
Gauge--ghost table vs. finite transformations & $2.398\times10^{-6}$ & Nonlinear ghost sector\\
Nonzero-background FP matrix & $3.234\times10^{-10}$ abs.; $1.679\times10^{-11}$ rel. & Background ghost kernel\\
\bottomrule
\end{tabular}
\caption{Selected local vertex-module audits. The tests cover the stated
components and routings; they do not sample every possible external momentum
or test a full one-loop Ward--Slavnov--Taylor identity.}
\label{tab:vertex-audit}
\end{table*}

The small discrepancies demonstrate consistency of the local derivatives
and discrete conventions within the audited tests. The nonlinear vertex
generators are not yet consumed by the complete matching integrator or by a
production gauge-sector determinant. We therefore distinguish
``implementation validated'' from ``physical matching completed''
throughout this paper.

\section{Matching definition and numerical projections}
\label{sec:matching-method}

\subsection{Amplitude matching and operator projections}
\label{subsec:amplitude-matching}

For a given operator, its coefficient is determined by projecting the
continuum-renormalized and lattice 1PI functions onto the same local
structure at identical external momenta. For the Higgs kinetic term,
\begin{equation}
 \delta Z_H(\mu_R)=
 \left.\frac{\partial\Sigma^{(2)}_{\MSbar,\rm ren}}
 {\partial p^2}\right|_{p^2=\mu_R^2}
 -\left.\frac{\partial\Sigma^{(2)}_{\rm lat}}
 {\partial\hat p^2}\right|_{\hat p^2=\mu_R^2}.
 \label{eq:z-projector}
\end{equation}
The lattice derivative is with respect to the lattice kernel variable
$\hat p^2$, not the continuum $p^2$. In the numerical implementation, the
two-point function is fit to a quadratic in its appropriate kernel
coordinate; the linear coefficient at the subtraction point is retained.
This projection separates the kinetic term from mass and higher-derivative
contributions over the selected fit range. The design matrix represents
$c_0+c_1x+c_2x^2$, with $x=p^2$ in the continuum and $x=\hat p^2$ on the
lattice. Its unit-slope test agrees with unity to $1.11\times10^{-15}$;
this is an algebraic projector test, not a loop-level Ward test.

For the scalar potential, a zero-momentum projector is valid in the
scalar-only calculation because the internal scalar lines are massive at
$\rho=0$. A contracted $2n$-point tensor in the gauge-linked calculation
uses nonzero, nonexceptional external momenta. The contracted amplitude is
fit as
\begin{equation}
 \mathcal{A}_{2n}(\bar s)=d_0+d_1\bar s+d_2\bar s^2,
 \qquad \delta c_{2n}\propto d_0,
 \label{eq:contact-projector}
\end{equation}
with separate continuum and lattice design matrices. The numerical
implementation contracts the full external-index tensor before fitting
the local intercept. Continuum and lattice fits use separate coordinates
because $p^2$ and $\hat p^2$ differ at finite $a$. This removes leading
momentum dependence within the chosen quadratic model, but does not
establish that the model captures all regulator dependence. The proportionality factor follows from the normalization of the $\rho^{2n}/(2n)$ operator and
the external-index contraction. Only when the full amplitude and the
appropriate normalization are established may this intercept be identified
with a matching coefficient.

For gauge two-point tensors, define, in either regularization,
\begin{equation}
 T_{\mu\nu}=K^2\delta_{\mu\nu}-K_\mu K_\nu,\qquad
 L_{\mu\nu}=K_\mu K_\nu,
 \label{eq:transverse-tensors}
\end{equation}
where $K=p$ in the continuum and $K_\mu=(2/a)\sin(ap_\mu/2)$ on the
lattice. The four-dimensional projectors are
\begin{equation}
 P_T^{\mu\nu}=\frac{T^{\mu\nu}}{3(K^2)^2},\qquad
 P_L^{\mu\nu}=\frac{K^\mu K^\nu}{(K^2)^2}.
 \label{eq:gauge-projectors}
\end{equation}
They isolate transverse and longitudinal form factors when the tensor has
the assumed decomposition. A residual longitudinal component is a useful
diagnostic, but its interpretation depends on the gauge fixing and the
infrared regulator.

\subsection{Fixed routes and common infrared prescription}
\label{subsec:routing}

The two-point external momenta are $p=\sqrt{s}\,\hat n$ with
$\hat n=(1,2,3,4)/\sqrt{30}$ and $s\in\{0.2,0.4,0.6\}$. Thus the
subtraction-point external momentum has $p^2=\mu_R^2=0.4$, while the
corresponding lattice coordinate is $\hat p^2=0.3966567139724254$. The
asymmetric $H^\dagger H A$ triangle uses
\begin{equation}
 \begin{aligned}
 p_{H^\dagger}&=\sqrt{0.4}(1,0,0,0),\\
 p_H&=\sqrt{0.4}(0.2,0.4,0.8,0),\\
 p_A&=-p_{H^\dagger}-p_H.
 \end{aligned}
 \label{eq:triangle-route}
\end{equation}
so that the three squared momenta are $0.4$, $0.336$, and $0.896$. Local
$2n$-scalar contacts use fixed-shape zero-sum routings at
$\bar s\in\{0.2,0.4,0.6\}$; no proper subset of external momenta sums to
zero. Nonexceptional routing avoids introducing spurious infrared
singularities into the contact projections.

Massless gauge and ghost propagators are regulated in both theories by
\begin{equation}
 \begin{aligned}
 k^2&\longmapsto k^2+\lambda_{\rm IR}^2,\\
 \hat k^2&\longmapsto\hat k^2+\lambda_{\rm IR}^2,\\
 \frac{\lambda_{\rm IR}}{\mu_R}
 &\in\left\{\frac18,\frac1{16},\frac1{32},\right.\\
 &\qquad\left.\frac1{64},\frac1{128},\frac1{256}\right\}.
 \end{aligned}
 \label{eq:ir-regulator}
\end{equation}
The limit $\lambda_{\rm IR}\to0$ is estimated from fits in
$x=(\lambda_{\rm IR}/\mu_R)^2$. The regulator is applied commonly to the
lattice and continuum integrands to make their difference more stable. It
is an auxiliary regulator, not a BRST-preserving mass term; consequently,
finite-regulator longitudinal residuals cannot by themselves be interpreted
as a continuum Ward-identity violation.

\subsection{Numerical integration and uncertainty accounting}
\label{subsec:qmc}

Lattice Brillouin-zone integrals are estimated with scrambled Sobol
sequences, and continuum and lattice integrands are evaluated on paired
points where the design permits. Scrambling and randomized quasi--Monte
Carlo provide independent replicates for estimating integration error
\cite{Sobol,Owen,SciPy}. Paired integration reduces noise in a difference when
the two integrands fluctuate similarly. The reported ``QMC SE'' is the
standard error across these replicates only.

Three uncertainty classes must be kept distinct:
\begin{enumerate}
\item \emph{Integration error:} estimated from independent scrambled
  replicates and quoted for the pilot amplitudes.
\item \emph{Regulator and fit-model error:} sensitivity to the range of
  $\lambda_{\rm IR}$ values and to the assumed extrapolation basis. This has
  not been quantified for any gauge-linked sector.
\item \emph{Truncation and omitted-sector error:} missing diagrams,
  higher-momentum operators, coupling matching, lattice-spacing effects,
  and the incomplete gauge/ghost determinant. These are not part of the
  reported QMC error.
\end{enumerate}
In particular, a small QMC SE does not certify a matching coefficient if
the fit is based on too few regulator values, held-out residuals are
coherent, or the relevant Ward identities have not been checked.

\section{Results}
\label{sec:results}

\subsection{Scalar-only radial matching and closure}
\label{subsec:scalar-results}

The scalar-only one-loop effective potential gives a finite, independently
reproducible scope in which gauge, link, and ghost loops are excluded. Set
$z=\rho^2$ and $D=\hat k^2$. The radial and transverse shifts from the
false-vacuum mass are
\begin{equation}
 P_L(z)=3\lambda z+5\gamma z^2+7\delta z^3,\qquad
 P_T(z)=\lambda z+\gamma z^2+\delta z^3.
 \label{eq:scalar-shifts}
\end{equation}
The one-loop lattice potential is
\begin{equation}
 \begin{aligned}
 V_{1,\rm lat}^{\rm scalar}(z)
 &=\frac12\int_{\BZ}\frac{\dd^4k}{(2\pi)^4}\\
 &\quad\times\bigl[\log(D+m_0^2+P_L)\\
 &\qquad+3\log(D+m_0^2+P_T)\bigr].
 \end{aligned}
 \label{eq:scalar-one-loop}
\end{equation}
The factor of three counts the Goldstone directions. Expanding the logarithms
in powers of $z$ and subtracting the corresponding continuum-renormalized
moments yields the local potential coefficients through $\rho^{12}$; the
derivation is given below.

\paragraph{Moment expansion and coefficient normalization.}
For the scalar-only expansion, define the lattice moments
\begin{equation}
 I_n^{\rm lat}=\int_{\BZ}\frac{\dd^4k}{(2\pi)^4}
 \frac{1}{(\hat k^2+m_0^2)^n}.
 \label{eq:lat-moments}
\end{equation}
Using a Schwinger parameter and the four-dimensional hypercubic lattice
kernel gives
\begin{equation}
 I_n^{\rm lat}=\frac{a^{2n-4}}{(n-1)!}
 \int_0^\infty\dd t\,t^{n-1}e^{-a^2m_0^2t}
 \left[e^{-2t}I_0(2t)\right]^4,
 \label{eq:lat-moment-bessel}
\end{equation}
where $I_0$ is the modified Bessel function. This Bessel-function
representation is standard in analyses of the lattice scalar propagator
\cite{PaladiniSexton}. The continuum moments in the chosen $\MSbar$
convention, obtained by dimensional regularization and modified minimal
subtraction, are \cite{tHooft1973,BBDM}:
\begin{align}
 I_1^{\MSbar}&=\frac{m_0^2}{16\pi^2}
 \left[\log\frac{m_0^2}{\mu_R^2}-1\right],\\
 I_2^{\MSbar}&=-\frac{1}{16\pi^2}
 \log\frac{m_0^2}{\mu_R^2},\\
 I_n^{\MSbar}&=\frac{\Gamma(n-2)}{16\pi^2\Gamma(n)}
 (m_0^2)^{2-n},\qquad n\geq3.
 \label{eq:cont-moments}
\end{align}
The pole-containing low moments are understood after $\MSbar$ subtraction.
At the benchmark $m_0^2=\mu_R^2$, the finite $I_2^{\MSbar}$ value is zero.

Expand the two logarithms in Eq.~\eqref{eq:scalar-one-loop} in powers of
$P_L$ and $P_T$, following the standard one-loop effective-potential
expansion \cite{CW,Jackiw}. If $[z^n]f(z)$ denotes the coefficient of
$z^n$, the coefficient $\Delta v_n$ in the potential difference is
\begin{equation}
 \begin{aligned}
 \Delta v_n
 &=\frac12\sum_{j=1}^{n}\frac{(-1)^{j+1}}{j}
   \left(I_j^{\MSbar}-I_j^{\rm lat}\right)\\
 &\quad\times [z^n]\left(P_L^j+3P_T^j\right),\\
 \delta c_{2n}&=2n\Delta v_n.
 \end{aligned}
 \label{eq:scalar-coefficients}
\end{equation}
For $n=1,\ldots,6$, the coefficients $\delta c_{2n}$ are respectively
$\delta m^2$, $\delta\lambda$, $\delta\gamma$, $\delta\delta$,
$\delta c_{10}$, and $\delta c_{12}$. The first two provide useful
checks:
\begin{align}
 \delta m^2&=6\lambda\left(I_1^{\MSbar}-I_1^{\rm lat}\right),
 \label{eq:mass-check}\\
 \delta\lambda&=16\gamma\left(I_1^{\MSbar}-I_1^{\rm lat}\right)
 -12\lambda^2\left(I_2^{\MSbar}-I_2^{\rm lat}\right).
 \label{eq:quartic-check}
\end{align}
The mass relation reproduces the $O(4)$ quartic-tadpole factor
$(N+2)\lambda=6\lambda$. The quartic receives both the single-insertion
sextic term and the two-vertex quartic ring \cite{Jackiw}. These relations are scalar-only
checks in the one-loop effective-potential expansion; they do not replace
the full-gauge amplitude projectors.

The resulting coefficients are listed in Table~\ref{tab:scalar-matching}.
The quadrature was repeated at tighter tolerances, changing each coefficient by
less than $1.1\times10^{-15}$; this is a quadrature-stability check rather
than an estimate of perturbative truncation error. At this order the scalar
two-point tadpole is momentum independent, so
$\delta Z_H^{\rm scalar}=0$.

\begin{table*}[t]
\centering
\small
\begin{tabular}{@{}rcccr@{}}
\toprule
$n$ & $I_n^{\rm lat}$ & $I_n^{\MSbar}$ & Coefficient & Value\\
\midrule
1 & $0.2268778187404072$ & $-0.002533029591058444$ & $\delta m^2$ & $2.75293017998$\\
2 & $0.03094303011979438$ & $0$ & $\delta\lambda$ & $-1.45119341289$\\
3 & $0.009469337125601956$ & $0.007915717472057639$ & $\delta\gamma$ & $-2.85119512150$\\
4 & $0.007100564330706442$ & $0.006596431226714698$ & $\delta\delta$ & $0.356724233481$\\
5 & $0.008539339433655424$ & $0.008245539033393373$ & $\delta c_{10}$ & $1.07431690553$\\
6 & $0.01264685637877819$ & $0.01236830855009006$ & $\delta c_{12}$ & $6.20491819587$\\
\bottomrule
\end{tabular}
\caption{Scalar-only radial matching in solver units at
$a=0.8$, $m_0^2=\mu_R^2=0.4$, $\lambda=-2$, $\gamma=0.8$, and
$\delta=0.08$. The sign is continuum-renormalized minus lattice. The
coefficients are not full-$\SUtwo$ matching results and must not be
inserted into the full-theory coefficient template.}
\label{tab:scalar-matching}
\end{table*}

The need for $\rho^{10}$ and $\rho^{12}$ is structural, not a numerical
artifact. The scalar determinant contains powers of $P_L$ and $P_T$ beyond
 the tree-level $\rho^8$ term; extracting its local expansion through the
same loop order generates additional operators. The reported values are
finite differences at the stated spacing and subtraction scale and are not
universal constants. They also do not close the full gauge theory, whose
massless gauge contribution is nonanalytic around $\rho=0$.

\subsection{Higgs two-point slope}
\label{subsec:higgs-slope}

\begin{figure*}[t]
\centering
\begin{tikzpicture}[scale=0.9]
  \begin{scope}[xshift=0cm]
    \draw[scalarline] (-1.35,0) -- (0,0);
    \draw[scalarline] (0,0) -- (1.35,0);
    \fill (0,0) circle (2pt);
    \draw[scalarline] (0,0) -- (0,0.35);
    \draw[scalarline] (0,0.95) circle (0.6);
    \node at (0,-0.55) {scalar tadpole};
    \node at (0,-0.9) {mass only};
  \end{scope}
  \begin{scope}[xshift=4.5cm]
    \draw[scalarline] (-1.5,0) -- (-1,0);
    \draw[scalarline] (1,0) -- (1.5,0);
    \draw[gaugelink] (-1,0) arc (180:0:1);
    \draw[goldline] (-1,0) arc (180:360:1);
    \fill (-1,0) circle (2pt) (1,0) circle (2pt);
    \node at (0,0.45) {$A$};
    \node at (0,-0.48) {$\chi$};
    \node at (0,-0.95) {$hA\chi$ exchange};
  \end{scope}
  \begin{scope}[xshift=9cm]
    \draw[scalarline] (-1.45,0) -- (1.45,0);
    \draw[gaugelink] (0,0.4) circle (0.6);
    \fill (0,0) circle (2pt);
    \node at (0,-0.65) {$hhAA$ seagull};
  \end{scope}
  \begin{scope}[xshift=13.4cm]
    \draw[scalarline] (-1.45,0) -- (1.45,0);
    \draw[ghostline] (-0.85,0) arc (180:0:0.85);
    \draw[ghostline] (0.85,0) arc (0:-180:0.85);
    \fill (-0.85,0) circle (2pt) (0.85,0) circle (2pt);
    \draw[red,very thick] (-0.32,-0.32) -- (0.32,0.32);
    \draw[red,very thick] (-0.32,0.32) -- (0.32,-0.32);
    \node at (0,-0.95) {direct ghost loop vanishes};
  \end{scope}
\end{tikzpicture}
\caption{Selected Higgs two-point diagrams at $\bar\rho=0$.
Gauge exchange and the link seagull determine the slope difference; the
scalar tadpole is momentum independent. The direct ghost--Higgs diagram
vanishes because the Faddeev--Popov operator is independent of the Higgs
background at this point.}
\label{fig:higgs-diagrams}
\end{figure*}
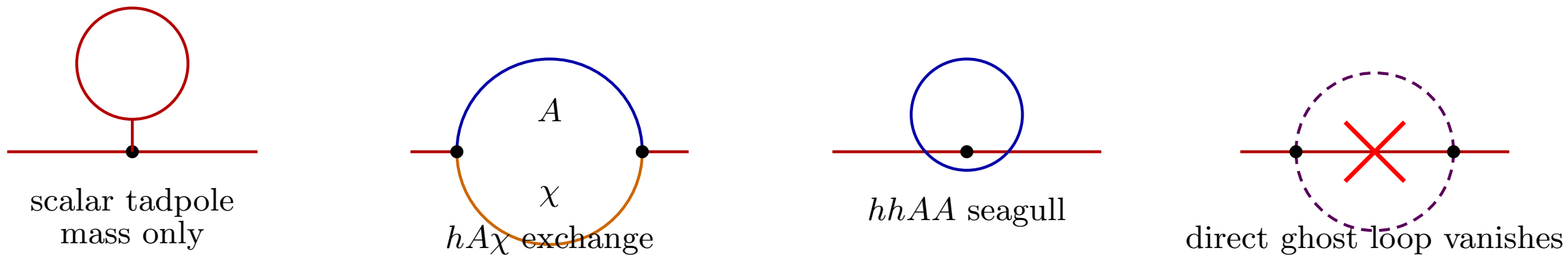

At the symmetric background, the one-loop Higgs two-point diagrams in the
selected gauge are the scalar quartic tadpole, gauge exchange through the
$hA\chi$ vertex, and the $hhAA$ seagull. The Faddeev--Popov operator has no
direct Higgs coupling at $\bar\rho=0$ in this setup. The scalar tadpole
contributes to the mass but not the slope. Figure~\ref{fig:higgs-diagrams}
shows these diagram families and the vanishing direct ghost--Higgs loop.

The regulator sequence and the extrapolated result are shown in
Table~\ref{tab:zh-results}. Three points determine the quadratic fit in
$x=(\lambda_{\rm IR}/\mu_R)^2$ exactly; thus the quoted intercept has no
independent fourth-point test. The monotone sequence and small replicate
error support numerical stability of this particular combination, but the
reported uncertainty is still QMC-only.

\begin{table}[t]
\centering
\small
\begin{tabular}{@{}ccc@{}}
\toprule
$\lambda_{\rm IR}/\mu_R$ & $\delta Z_H^{\MSbar-\rm lat}$ & QMC SE\\
\midrule
$1/8$ & $0.0328817051$ & $3.41\times10^{-7}$\\
$1/16$ & $0.0329036915$ & $3.45\times10^{-7}$\\
$1/32$ & $0.0329097345$ & $3.55\times10^{-7}$\\
$0$ (quadratic fit) & $0.0329117974$ & $3.60\times10^{-7}$\\
\bottomrule
\end{tabular}
\caption{Higgs kinetic matching projection at the benchmark. The last row
is the $\lambda_{\rm IR}\to0$ fit intercept.}
\label{tab:zh-results}
\end{table}

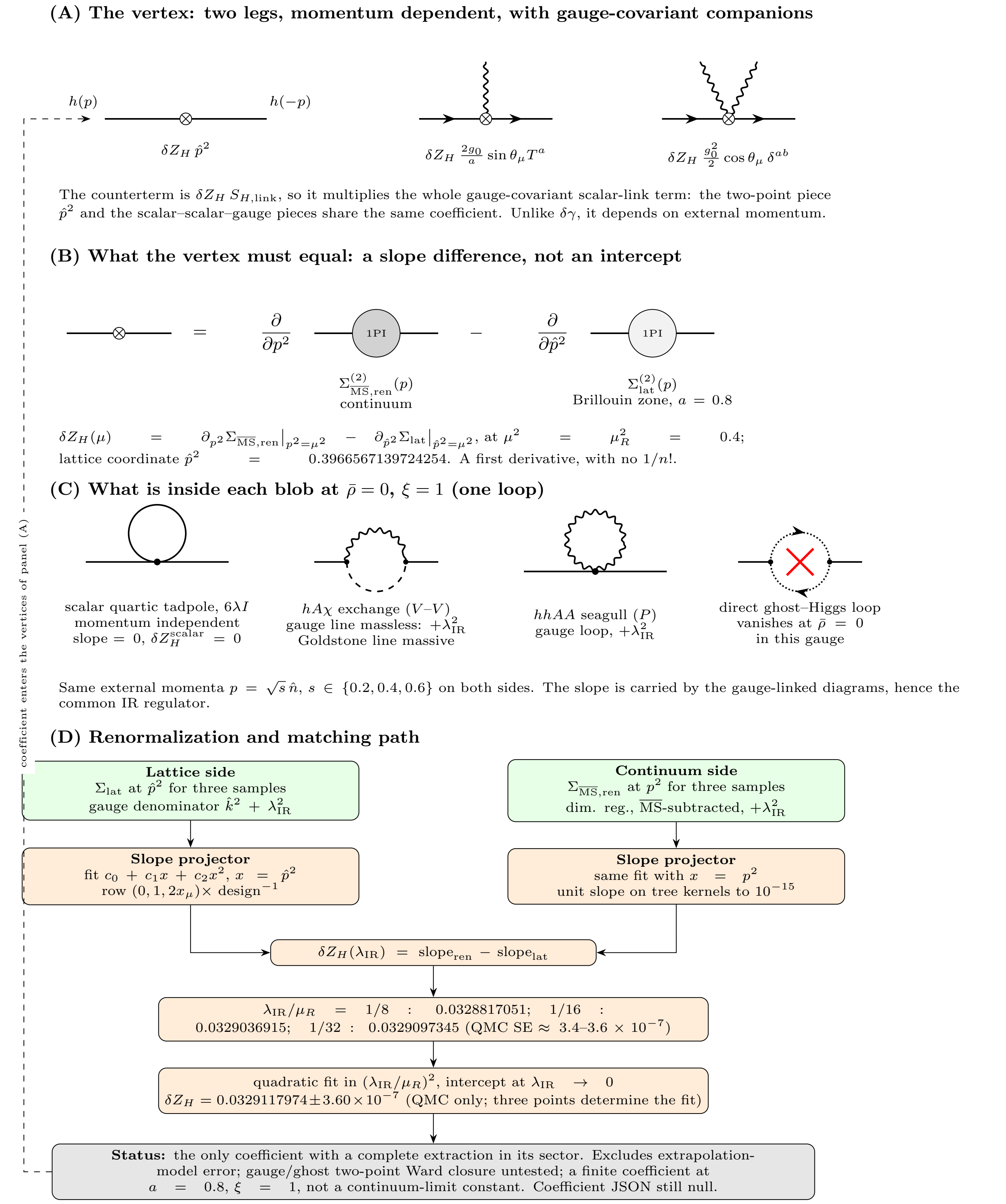
\begin{figure*}[p]
\centering
\resizebox{0.98\textwidth}{!}{%
\begin{tikzpicture}[scale=0.85]
\node[anchor=west,font=\small\bfseries] at (0,24.2) {(A) The vertex: two legs, momentum dependent, with gauge-covariant companions};
\begin{scope}[shift={(3.0,22.0)}]
  \draw[higgs] (-1.7,0) -- (1.7,0); \ctv{0,0}
  \node[font=\scriptsize] at (-2.15,0.35) {$h(p)$}; \node[font=\scriptsize] at (2.2,0.35) {$h(-p)$};
  \node[font=\scriptsize] at (0,-0.65) {$\delta Z_H\,\hat p^2$};
\end{scope}
\begin{scope}[shift={(9.3,22.0)}]
  \draw[scal] (-1.4,0) -- (0,0); \draw[scal] (0,0) -- (1.4,0); \draw[gauge] (0,0) -- (0,1.2); \ctv{0,0}
  \node[font=\scriptsize] at (0,-0.75) {$\delta Z_H\,\frac{2g_0}{a}\sin\theta_\mu T^a$};
\end{scope}
\begin{scope}[shift={(14.4,22.0)}]
  \draw[scal] (-1.4,0) -- (0,0); \draw[scal] (0,0) -- (1.4,0);
  \draw[gauge] (0,0) -- (-0.6,1.2); \draw[gauge] (0,0) -- (0.6,1.2); \ctv{0,0}
  \node[font=\scriptsize] at (0,-0.75) {$\delta Z_H\,\frac{g_0^2}{2}\cos\theta_\mu\,\delta^{ab}$};
\end{scope}
\node[anchor=west,align=left,font=\scriptsize,text width=16.5cm] at (0.2,20.2)
 {The counterterm is $\delta Z_H\,S_{H,\rm link}$, so it multiplies the whole gauge-covariant scalar-link term: the two-point piece $\hat p^2$ and the scalar--scalar--gauge pieces share the same coefficient. Unlike $\delta\gamma$, it depends on external momentum.};
\node[anchor=west,font=\small\bfseries] at (0,19.1) {(B) What the vertex must equal: a slope difference, not an intercept};
\begin{scope}[shift={(1.6,17.5)}]
  \draw[higgs] (-1.1,0) -- (1.1,0); \ctv{0,0}
\end{scope}
\node at (3.3,17.5) {$=$};
\node at (4.9,17.5) {$\dfrac{\partial}{\partial p^2}$};
\begin{scope}[shift={(7.0,17.5)}]
  \draw[higgs] (-1.3,0) -- (1.3,0); \draw[fill=gray!35] (0,0) circle (0.5); \node[font=\tiny] at (0,0) {1PI};
\end{scope}
\node at (9.1,17.5) {$-$};
\node at (10.7,17.5) {$\dfrac{\partial}{\partial \hat p^2}$};
\begin{scope}[shift={(12.8,17.5)}]
  \draw[higgs] (-1.3,0) -- (1.3,0); \draw[fill=gray!10] (0,0) circle (0.5); \node[font=\tiny] at (0,0) {1PI};
\end{scope}
\node[font=\scriptsize,align=center,text width=3.4cm] at (7.0,16.3) {$\Sigma^{(2)}_{\overline{\rm MS},\rm ren}(p)$\\continuum};
\node[font=\scriptsize,align=center,text width=3.4cm] at (12.8,16.3) {$\Sigma^{(2)}_{\rm lat}(p)$\\Brillouin zone, $a=0.8$};
\node[anchor=west,align=left,font=\scriptsize,text width=16.5cm] at (0.2,15.1)
 {$\delta Z_H(\mu)=\partial_{p^2}\Sigma_{\overline{\rm MS},\rm ren}\big|_{p^2=\mu^2}-\partial_{\hat p^2}\Sigma_{\rm lat}\big|_{\hat p^2=\mu^2}$, at $\mu^2=\mu_R^2=0.4$; lattice coordinate $\hat p^2=0.3966567139724254$. A first derivative, with no $1/n!$.};
\node[anchor=west,font=\small\bfseries] at (0,14.2) {(C) What is inside each blob at $\bar\rho=0$, $\xi=1$ (one loop)};
\begin{scope}[shift={(2.4,12.7)}]
  \draw[higgs] (-1.5,0) -- (1.5,0); \draw[higgs] (0,0.6) circle (0.6); \fill (0,0) circle (2pt);
\end{scope}
\node[align=center,font=\scriptsize,text width=3.9cm] at (2.4,11.4) {scalar quartic tadpole, $6\lambda I$\\momentum independent\\slope $=0$, $\delta Z_H^{\rm scalar}=0$};
\begin{scope}[shift={(7.0,12.7)},scale=0.62]
  \draw[higgs] (-2.1,0) -- (-1,0); \draw[higgs] (1,0) -- (2.1,0);
  \draw[gauge] (-1,0) arc (180:0:1); \draw[gold] (-1,0) arc (180:360:1);
  \fill (-1,0) circle (2.5pt) (1,0) circle (2.5pt);
\end{scope}
\node[align=center,font=\scriptsize,text width=3.9cm] at (7.0,11.4) {$hA\chi$ exchange ($V$--$V$)\\gauge line massless: $+\lambda_{\rm IR}^2$\\Goldstone line massive};
\begin{scope}[shift={(11.6,12.5)}]
  \draw[higgs] (-1.5,0) -- (1.5,0); \draw[gauge] (0,0.6) circle (0.6); \fill (0,0) circle (2pt);
\end{scope}
\node[align=center,font=\scriptsize,text width=3.9cm] at (11.6,11.4) {$hhAA$ seagull ($P$)\\gauge loop, $+\lambda_{\rm IR}^2$};
\begin{scope}[shift={(15.9,12.7)},scale=0.62]
  \draw[higgs] (-2.1,0) -- (-1,0); \draw[higgs] (1,0) -- (2.1,0);
  \draw[ghost] (-1,0) arc (180:0:1); \draw[ghost] (1,0) arc (0:-180:1);
  \fill (-1,0) circle (2.5pt) (1,0) circle (2.5pt);
  \draw[red,very thick] (-0.45,-0.45) -- (0.45,0.45); \draw[red,very thick] (-0.45,0.45) -- (0.45,-0.45);
\end{scope}
\node[align=center,font=\scriptsize,text width=3.4cm] at (15.9,11.4) {direct ghost--Higgs loop\\vanishes at $\bar\rho=0$ in this gauge};
\node[anchor=west,align=left,font=\scriptsize,text width=16.5cm] at (0.2,9.9)
 {Same external momenta $p=\sqrt s\,\hat n$, $s\in\{0.2,0.4,0.6\}$ on both sides. The slope is carried by the gauge-linked diagrams, hence the common IR regulator.};
\node[anchor=west,font=\small\bfseries] at (0,9.0) {(D) Renormalization and matching path};
\pbox{fill=green!10,text width=5.8cm} (lat) at (3.1,7.9) {\textbf{Lattice side}\\$\Sigma_{\rm lat}$ at $\hat p^2$ for three samples\\gauge denominator $\hat k^2+\lambda_{\rm IR}^2$};
\pbox{fill=green!10,text width=5.8cm} (dr) at (13.3,7.9) {\textbf{Continuum side}\\$\Sigma_{\overline{\rm MS},\rm ren}$ at $p^2$ for three samples\\dim. reg., $\overline{\rm MS}$-subtracted, $+\lambda_{\rm IR}^2$};
\pbox{fill=orange!15,text width=5.8cm} (sl) at (3.1,6.1) {\textbf{Slope projector}\\fit $c_0+c_1x+c_2x^2$, $x=\hat p^2$\\row $(0,1,2x_\mu)\times$ design$^{-1}$};
\pbox{fill=orange!15,text width=5.8cm} (sc) at (13.3,6.1) {\textbf{Slope projector}\\same fit with $x=p^2$\\unit slope on tree kernels to $10^{-15}$};
\pbox{fill=orange!15,text width=5.6cm} (df) at (8.2,4.5) {$\delta Z_H(\lambda_{\rm IR})=\text{slope}_{\rm ren}-\text{slope}_{\rm lat}$};
\pbox{fill=orange!15,text width=9.6cm} (tb) at (8.2,3.1) {$\lambda_{\rm IR}/\mu_R=1/8:\ 0.0328817051;\quad 1/16:\ 0.0329036915;\quad 1/32:\ 0.0329097345$ (QMC SE $\approx3.4$--$3.6\times10^{-7}$)};
\pbox{fill=orange!15,text width=9.6cm} (fi) at (8.2,1.6) {quadratic fit in $(\lambda_{\rm IR}/\mu_R)^2$, intercept at $\lambda_{\rm IR}\to0$\\$\delta Z_H=0.0329117974\pm3.60\times10^{-7}$ (QMC only; three points determine the fit)};
\pbox{fill=gray!20,text width=13.4cm} (out) at (8.2,-0.1) {\textbf{Status:} the only coefficient with a complete extraction in its sector. Excludes extrapolation-model error; gauge/ghost two-point Ward closure untested; a finite coefficient at $a=0.8$, $\xi=1$, not a continuum-limit constant. Coefficient JSON still null.};
\draw[-Stealth] (lat) -- (sl); \draw[-Stealth] (dr) -- (sc);
\draw[-Stealth] (sl.south) -- (3.1,4.5) -- (df.west);
\draw[-Stealth] (sc.south) -- (13.3,4.5) -- (df.east);
\draw[-Stealth] (df) -- (tb); \draw[-Stealth] (tb) -- (fi); \draw[-Stealth] (fi) -- (out);
\draw[-Stealth,dashed] (out.west) -- (-0.4,-0.1) -- (-0.4,22.0) -- (1.0,22.0);
\node[font=\tiny,rotate=90,fill=white] at (-0.4,11) {coefficient enters the vertices of panel (A)};
\end{tikzpicture}%
}
\caption{Anatomy of the kinetic counterterm $\delta Z_H$ for contrast with
the six-point potential coefficient: (A) the momentum-dependent two-leg
vertex and its gauge-covariant companions; (B) the slope identity; (C) the
one-loop families at $\bar\rho=0$; (D) the path to the extracted value.
Topologies are representative.}
\label{fig:zh_anatomy}
\end{figure*}

Figure~\ref{fig:zh_anatomy} summarizes how this slope projection defines
$\delta Z_H$, including the gauge-covariant counterterm vertices and the
one-loop diagrams contributing at the symmetric background.

\subsection{Gauge-linked local contact pilots}
\label{subsec:gauge-pilots}

\begin{figure}[!t]
\centering
\begin{tikzpicture}
\begin{axis}[
  width=0.98\columnwidth,
  height=5.0cm,
  xlabel={$\lambda_{\rm IR}/\mu_R$},
  ylabel={$\DeltaGamma^{(4)}_{\rm gauge}$},
  xmin=0,xmax=0.14,
  ymin=0.2530,ymax=0.2552,
  xtick={0,0.03125,0.0625,0.09375,0.125},
  xticklabels={0,$1/32$,$1/16$,$3/32$,$1/8$},
  tick label style={font=\scriptsize},
  label style={font=\small},
  legend style={font=\scriptsize,at={(0.02,0.98)},anchor=north west},
  grid=major,
  grid style={gray!20}
]
\addplot+[only marks,mark=*,blue!70!black,
  error bars/.cd,y dir=both,y explicit]
coordinates {
  (0.125,0.2542605) +- (0,0.000334)
  (0.0625,0.2541718) +- (0,0.000543)
  (0.03125,0.2540645) +- (0,0.000681)
  (0.015625,0.2539542) +- (0,0.000753)
  (0.0078125,0.2538960) +- (0,0.000783)
  (0.00390625,0.2538767) +- (0,0.000792)
};
\addlegendentry{QMC data}
\addplot[black,dashed,domain=0:0.13,samples=2]{0.2540212};
\addlegendentry{three-point intercept}
\end{axis}
\end{tikzpicture}
\caption{Gauge-linked four-Higgs projection versus the common infrared
regulator. Error bars are QMC replicate errors only. The dashed line marks
the reported three-point extrapolated intercept; it is not a regulator-fit
curve and does not display model uncertainty.}
\label{fig:fourpoint-plot}
\end{figure}
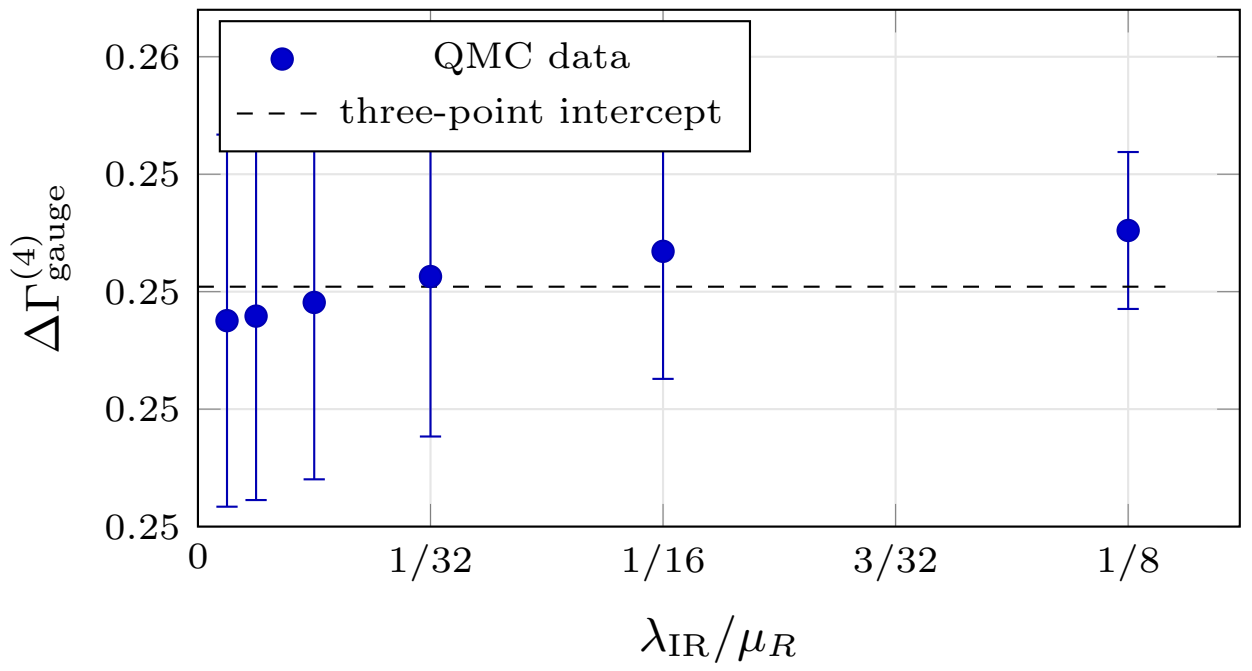

Gauge-linked four-Higgs amplitudes were integrated on six infrared-regulator
values. A quadratic fit in $x=(\lambda_{\rm IR}/\mu_R)^2$ to the first
three points gives the intercept
$\DeltaGamma^{(4)}_{\rm gauge}=0.2540212\pm0.0007361$, where the error is
the replicate QMC SE. The remaining three points were held out from that
fit. Their residuals have the same sign and are each approximately
$1.4$ paired standard errors below the three-point prediction. The data are
listed in Table~\ref{tab:fourpoint-regulator} and shown in
Fig.~\ref{fig:fourpoint-plot}.

\begin{table}[t]
\centering
\small
\begin{tabular}{@{}ccc@{}}
\toprule
$\lambda_{\rm IR}/\mu_R$ & $\DeltaGamma^{(4)}_{\rm gauge}$ & QMC SE\\
\midrule
$1/8$ & $0.2542605$ & $3.34\times10^{-4}$\\
$1/16$ & $0.2541718$ & $5.43\times10^{-4}$\\
$1/32$ & $0.2540645$ & $6.81\times10^{-4}$\\
$1/64$ & $0.2539542$ & $7.53\times10^{-4}$\\
$1/128$ & $0.2538960$ & $7.83\times10^{-4}$\\
$1/256$ & $0.2538767$ & $7.92\times10^{-4}$\\
\bottomrule
\end{tabular}
\caption{Gauge-linked four-Higgs projection. The first three regulator
values were used in the reported quadratic fit; the final three served as
held-out diagnostics. All quoted errors are QMC-only.}
\label{tab:fourpoint-regulator}
\end{table}

The gauge-linked eight-Higgs calculation used eight scrambled-Sobol
replicates, $2^{12}$ points per replicate, six regulators, and three fixed
routes, with continuum and lattice evaluations paired on common points.
Reaggregation of its checkpoint blocks reproduced the saved arrays to
roundoff. The fit-window results appear in Table~\ref{tab:eightpoint}. The five-point
intercept shifts by $-0.01347\pm0.00560$ relative to the three-point fit,
about $2.4$ paired SE. This does not establish a large physical effect, but
it is evidence that the replicate SE alone is not an adequate measure of
fit-window stability. The quantity remains a finite-contact projection,
not an accepted matching coefficient.

\begin{table}[t]
\centering
\small
\begin{tabular}{@{}lccc@{}}
\toprule
Fit window & Intercept & QMC SE & Paired shift\\
\midrule
3 regulators & $-19.85454$ & $1.30141$ & $0$\\
4 regulators & $-19.86357$ & $1.30335$ & $-0.00903\pm0.00373$\\
5 regulators & $-19.86800$ & $1.30421$ & $-0.01347\pm0.00560$\\
\bottomrule
\end{tabular}
\caption{Gauge-linked eight-Higgs fit-window comparison. The replicate SE
does not include regulator-model uncertainty.}
\label{tab:eightpoint}
\end{table}

The six-Higgs pilot was independently rerun in the November 12 pipeline
check. The earlier estimate, $-2.2646\pm0.0702$, and the rerun,
$-2.2199\pm0.1526$, differ by $0.27$ combined standard deviations, so the
two are statistically consistent at the precision of these runs. The rerun
used four scrambles and 64 points per scramble; it does not resolve the
regulator-fit question. In pipeline version 1.1, the stability gates were
classified as underpowered unless at least eight scrambles were available.
The scalar-only coefficient table was reproduced to $4.9\times10^{-12}$ in
that pipeline check, but this validates only the scalar-only stage. These
checks support computational provenance only. For public reproducibility,
archive the source, run manifests, raw replicate outputs, and analysis
scripts alongside any final coefficient table. No complete $\SUtwo$ coefficient file is claimed here.

Ten- and twelve-Higgs gauge-linked values were also explored, with reported
central values of order $-1.24\times10^4$ and $-2.17\times10^6$,
respectively. Each used only four points per scramble. We therefore classify
them as exploratory outputs and do not use them to infer coefficients or
the convergence of the operator expansion.

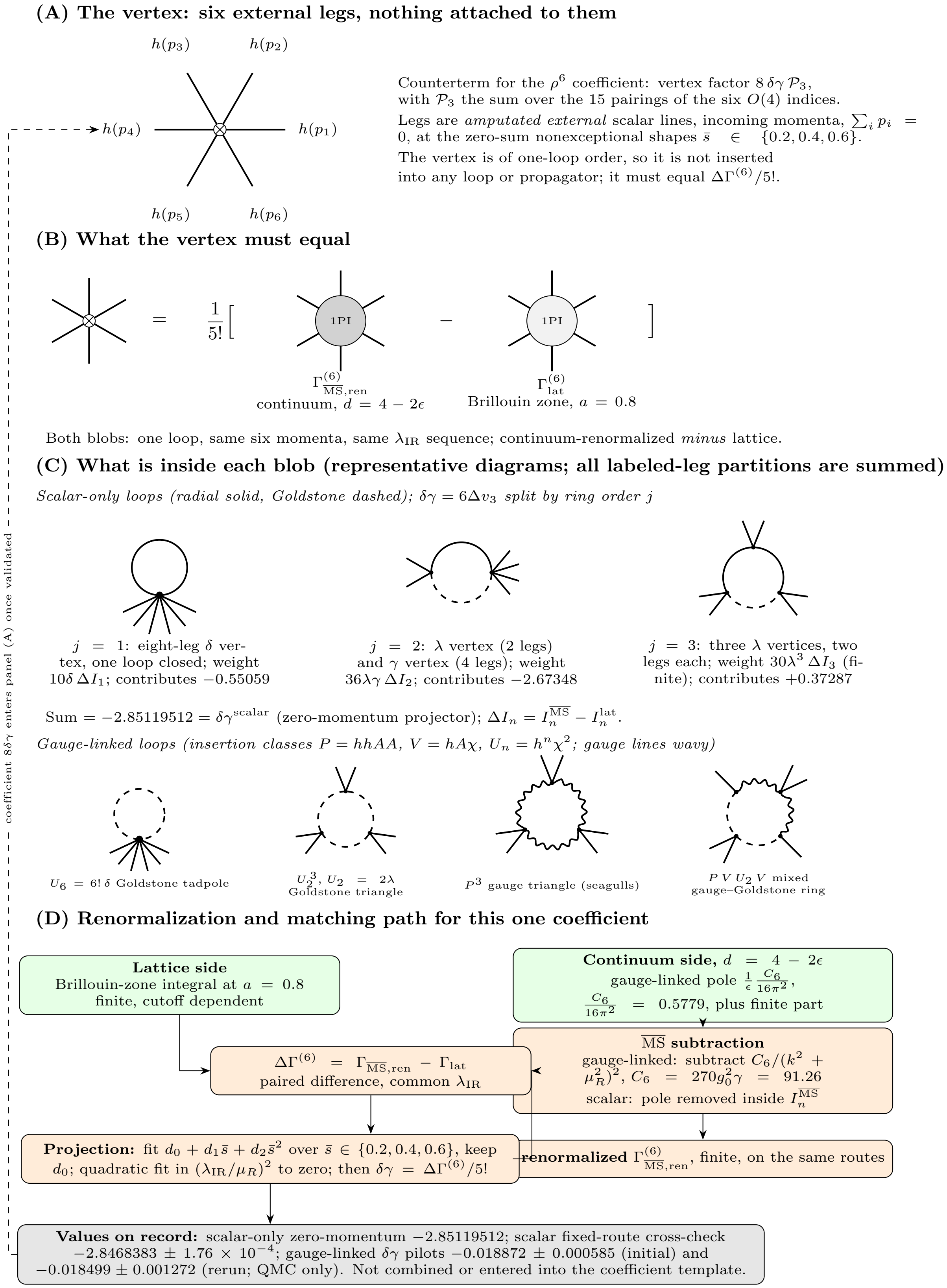
\begin{figure*}[p]
\centering
\resizebox{0.94\textwidth}{!}{%
\begin{tikzpicture}[scale=0.85]
\node[anchor=west,font=\small\bfseries] at (0,24.2) {(A) The vertex: six external legs, nothing attached to them};
\begin{scope}[shift={(3.8,21.9)}]
  \foreach \k in {1,...,6}{
    \draw[higgs] (0,0) -- ({60*(\k-1)}:1.3);
    \node[font=\scriptsize] at ({60*(\k-1)}:1.95) {$h(p_{\k})$};
  }
  \ctv{0,0}
\end{scope}
\node[anchor=west,align=left,font=\scriptsize,text width=9.6cm] at (7.2,21.9)
 {Counterterm for the $\rho^6$ coefficient: vertex factor $8\,\delta\gamma\,\mathcal P_3$, with $\mathcal P_3$ the sum over the $15$ pairings of the six $O(4)$ indices.\\[2pt]
  Legs are \emph{amputated external} scalar lines, incoming momenta, $\sum_i p_i=0$, at the zero-sum nonexceptional shapes $\bar s\in\{0.2,0.4,0.6\}$.\\[2pt]
  The vertex is of one-loop order, so it is not inserted into any loop or propagator; it must equal $\Delta\Gamma^{(6)}/5!$.};
\node[anchor=west,font=\small\bfseries] at (0,19.7) {(B) What the vertex must equal};
\begin{scope}[shift={(1.2,18.1)}]
  \foreach \k in {1,...,6}{\draw[higgs] (0,0) -- ({30+60*(\k-1)}:0.85);}
  \ctv{0,0}
\end{scope}
\node at (2.6,18.1) {$=$};
\node at (3.8,18.1) {$\dfrac1{5!}\Bigl[$};
\begin{scope}[shift={(6.2,18.1)}]
  \foreach \k in {1,...,6}{\draw[higgs] (0,0) -- ({30+60*(\k-1)}:1.0);}
  \draw[fill=gray!35] (0,0) circle (0.5); \node[font=\tiny] at (0,0) {1PI};
\end{scope}
\node at (8.3,18.1) {$-$};
\begin{scope}[shift={(10.4,18.1)}]
  \foreach \k in {1,...,6}{\draw[higgs] (0,0) -- ({30+60*(\k-1)}:1.0);}
  \draw[fill=gray!10] (0,0) circle (0.5); \node[font=\tiny] at (0,0) {1PI};
\end{scope}
\node at (12.4,18.1) {$\Bigr]$};
\node[align=center,font=\scriptsize,text width=3.4cm] at (6.2,16.7) {$\Gamma^{(6)}_{\overline{\rm MS},\rm ren}$\\continuum, $d=4-2\epsilon$};
\node[align=center,font=\scriptsize,text width=3.4cm] at (10.4,16.7) {$\Gamma^{(6)}_{\rm lat}$\\Brillouin zone, $a=0.8$};
\node[anchor=west,align=left,font=\scriptsize,text width=16cm] at (0.2,15.75)
 {Both blobs: one loop, same six momenta, same $\lambda_{\rm IR}$ sequence; continuum-renormalized \emph{minus} lattice.};
\node[anchor=west,font=\small\bfseries] at (0,15.2) {(C) What is inside each blob (representative diagrams; all labeled-leg partitions are summed)};
\node[anchor=west,font=\scriptsize\itshape] at (0,14.6) {Scalar-only loops (radial solid, Goldstone dashed); $\delta\gamma=6\Delta v_3$ split by ring order $j$};
\begin{scope}[shift={(2.6,13.1)}]
  \draw[higgs] (0,0.1) circle (0.55); \fill (0,-0.45) circle (2pt);
  \foreach \a in {205,230,255,285,310,335}{\draw[higgs] (0,-0.45) -- ++(\a:0.8);}
\end{scope}
\node[align=center,font=\scriptsize,text width=4.6cm] at (2.6,11.3) {$j=1$: eight-leg $\delta$ vertex, one loop closed; weight $10\delta\,\Delta I_1$; contributes $-0.55059$};
\begin{scope}[shift={(8.6,13.1)},scale=0.6]
  \arcE{higgs}{0}{180}\arcE{gold}{180}{360}\dotv{0}\dotv{180}
  \legn{180}{150}\legn{180}{210}\legn{0}{-50}\legn{0}{-17}\legn{0}{17}\legn{0}{50}
\end{scope}
\node[align=center,font=\scriptsize,text width=4.6cm] at (8.6,11.3) {$j=2$: $\lambda$ vertex (2 legs) and $\gamma$ vertex (4 legs); weight $36\lambda\gamma\,\Delta I_2$; contributes $-2.67348$};
\begin{scope}[shift={(14.4,13.0)},scale=0.6]
  \arcE{higgs}{90}{210}\arcE{gold}{210}{330}\arcE{higgs}{330}{450}
  \dotv{90}\dotv{210}\dotv{330}
  \legn{90}{68}\legn{90}{112}\legn{210}{188}\legn{210}{232}\legn{330}{308}\legn{330}{352}
\end{scope}
\node[align=center,font=\scriptsize,text width=4.6cm] at (14.4,11.3) {$j=3$: three $\lambda$ vertices, two legs each; weight $30\lambda^3\,\Delta I_3$ (finite); contributes $+0.37287$};
\node[anchor=west,font=\scriptsize] at (0.2,10.25) {Sum $=-2.85119512=\delta\gamma^{\rm scalar}$ (zero-momentum projector); $\Delta I_n=I_n^{\overline{\rm MS}}-I_n^{\rm lat}$.};
\node[anchor=west,font=\scriptsize\itshape] at (0,9.7) {Gauge-linked loops (insertion classes $P=hhAA$, $V=hA\chi$, $U_n=h^n\chi^2$; gauge lines wavy)};
\begin{scope}[shift={(2.2,8.2)}]
  \draw[gold] (0,0.1) circle (0.5); \fill (0,-0.4) circle (2pt);
  \foreach \a in {205,230,255,285,310,335}{\draw[higgs] (0,-0.4) -- ++(\a:0.65);}
\end{scope}
\node[align=center,font=\tiny,text width=3.4cm] at (2.2,6.9) {$U_6=6!\,\delta$ Goldstone tadpole};
\begin{scope}[shift={(6.3,8.2)},scale=0.55]
  \arcE{gold}{90}{210}\arcE{gold}{210}{330}\arcE{gold}{330}{450}
  \dotv{90}\dotv{210}\dotv{330}
  \legn{90}{68}\legn{90}{112}\legn{210}{188}\legn{210}{232}\legn{330}{308}\legn{330}{352}
\end{scope}
\node[align=center,font=\tiny,text width=3.4cm] at (6.3,6.9) {$U_2^{\,3}$, $U_2=2\lambda$ Goldstone triangle};
\begin{scope}[shift={(10.4,8.3)},scale=0.62]
  \arcE{gauge}{90}{210}\arcE{gauge}{210}{330}\arcE{gauge}{330}{450}
  \dotv{90}\dotv{210}\dotv{330}
  \legn{90}{68}\legn{90}{112}\legn{210}{188}\legn{210}{232}\legn{330}{308}\legn{330}{352}
\end{scope}
\node[align=center,font=\tiny,text width=3.4cm] at (10.4,6.9) {$P^3$ gauge triangle (seagulls)};
\begin{scope}[shift={(14.5,8.3)},scale=0.62]
  \arcE{gauge}{45}{135}\arcE{gold}{135}{225}\arcE{gold}{225}{315}\arcE{gauge}{315}{405}
  \dotv{45}\dotv{135}\dotv{225}\dotv{315}
  \legn{45}{25}\legn{45}{65}\legn{135}{135}\legn{225}{205}\legn{225}{245}\legn{315}{315}
\end{scope}
\node[align=center,font=\tiny,text width=3.4cm] at (14.5,6.9) {$P\,V\,U_2\,V$ mixed gauge--Goldstone ring};
\node[anchor=west,font=\small\bfseries] at (0,6.2) {(D) Renormalization and matching path for this one coefficient};
\pbox{fill=green!10,text width=5.2cm} (lat) at (3.0,4.9) {\textbf{Lattice side}\\Brillouin-zone integral at $a=0.8$\\finite, cutoff dependent};
\pbox{fill=green!10,text width=6.2cm} (dr) at (13.4,4.9) {\textbf{Continuum side, $d=4-2\epsilon$}\\gauge-linked pole $\frac1\epsilon\frac{C_6}{16\pi^2}$, $\frac{C_6}{16\pi^2}=0.5779$, plus finite part};
\pbox{fill=orange!15,text width=6.2cm} (ms) at (13.4,3.2) {\textbf{$\overline{\rm MS}$ subtraction}\\gauge-linked: subtract $C_6/(k^2+\mu_R^2)^2$, $C_6=270g_0^2\gamma=91.26$\\scalar: pole removed inside $I_n^{\overline{\rm MS}}$};
\pbox{fill=orange!15,text width=6.2cm} (ren) at (13.4,1.45) {\textbf{renormalized} $\Gamma^{(6)}_{\overline{\rm MS},\rm ren}$, finite, on the same routes};
\pbox{fill=orange!15,text width=5.2cm} (dg) at (6.8,3.2) {$\Delta\Gamma^{(6)}=\Gamma_{\overline{\rm MS},\rm ren}-\Gamma_{\rm lat}$\\paired difference, common $\lambda_{\rm IR}$};
\pbox{fill=orange!15,text width=8.2cm} (pr) at (4.8,1.45) {\textbf{Projection:} fit $d_0+d_1\bar s+d_2\bar s^2$ over $\bar s\in\{0.2,0.4,0.6\}$, keep $d_0$; quadratic fit in $(\lambda_{\rm IR}/\mu_R)^2$ to zero; then $\delta\gamma=\Delta\Gamma^{(6)}/5!$};
\pbox{fill=gray!20,text width=12.4cm} (out) at (7.2,-0.45) {\textbf{Values on record:} scalar-only zero-momentum $-2.85119512$; scalar fixed-route cross-check $-2.8468383\pm1.76\times10^{-4}$; gauge-linked $\delta\gamma$ pilots $-0.018872\pm0.000585$ (initial) and $-0.018499\pm0.001272$ (rerun; QMC only). Not combined or entered into the coefficient template.};
\draw[-Stealth] (dr) -- (ms); \draw[-Stealth] (ms) -- (ren);
\draw[-Stealth] (ren.west) -- (10.0,1.45) -- (10.0,3.2) -- (dg.east);
\draw[-Stealth] (lat.south) -- (3.0,3.2) -- (dg.west);
\draw[-Stealth] (dg) -- (pr.north -| dg.south); \draw[-Stealth] (pr) -- (pr |- out.north);
\draw[-Stealth,dashed] (out.west) -- (-0.4,-0.45) -- (-0.4,21.9) -- (1.4,21.9);
\node[font=\tiny,rotate=90,fill=white] at (-0.4,11) {coefficient $8\delta\gamma$ enters panel (A) once validated};
\end{tikzpicture}%
}
\caption{Anatomy of the six-leg counterterm vertex ($\delta\gamma$, $\rho^6$ coefficient) of Eq.~\ref{eq:potential-ct}: (A) vertex; (B) identity it implements; (C) one-loop families inside the two blobs; (D) renormalization and matching path. Topologies are representative; the split by ring order is arithmetic on the tabulated moments and sums to the scalar-only zero-momentum coefficient in Table~\ref{tab:scalar-matching}.}
\label{fig:gamma_anatomy}
\end{figure*}

The six-leg normalization and the separation between the scalar-only result and
the gauge-linked pilot estimates are summarized in Fig.~\ref{fig:gamma_anatomy}.
The gauge-linked values remain preliminary and are not combined with the
scalar-only coefficient.

\subsection{Gauge two-point diagnostics and identity status}
\label{subsec:gauge-two-point}

The gauge two-point function requires matter, pure-gauge, ghost, seagull,
and Haar-measure contributions, followed by a continuum subtraction and
transverse/longitudinal projection. The scalar-bubble plus seagull
matter-only check passes in the unregulated massive-line test. The continuum
one-loop tensor at the symmetric background is now derived and internally
validated in Sec.~\ref{sec:g2pt}; the raw lattice tensor has not yet been
compared with it, and the full identity closure remains open.

In the raw lattice follow-up that included the logarithmic-link Haar term,
the maximum normalized transverse residual was $0.4123$. At
$p^2=0.4$ and $\lambda_{\rm IR}/\mu_R=1/32$, one projected Ward component
was $-0.01198\pm0.00052$, or $23.1$ QMC SE from zero. This number is not a
Ward-identity failure of the continuum theory: the auxiliary infrared mass
breaks BRST symmetry at finite regulator, and the full continuum tensor has
not yet been subtracted from the lattice result. It is instead a warning
that the raw lattice sector needs a controlled common-regulator
extrapolation and a convention-matched comparison before any identity-based
conclusion is possible. The continuum calculation in Sec.~\ref{sec:g2pt}
quantifies its own regulator-induced longitudinal term, but it does not by
itself establish that the lattice and continuum normalizations agree.

\begin{figure*}[!t]
\centering
\begin{tikzpicture}[scale=0.82]
  \begin{scope}[xshift=0cm]
    \draw[gaugelink] (-1.45,0) -- (-1,0);
    \draw[gaugelink] (1,0) -- (1.45,0);
    \draw[scalarline] (-1,0) arc (180:0:1);
    \draw[scalarline] (1,0) arc (0:-180:1);
    \fill (-1,0) circle (2pt) (1,0) circle (2pt);
    \node at (0,-1.0) {matter bubble};
  \end{scope}
  \begin{scope}[xshift=4.5cm]
    \draw[gaugelink] (-1.45,0) -- (1.45,0);
    \draw[scalarline] (0,0.45) circle (0.6);
    \fill (0,0) circle (2pt);
    \node at (0,-1.0) {$AA\phi\phi$ seagull};
  \end{scope}
  \begin{scope}[xshift=9cm]
    \draw[gaugelink] (-1.45,0) -- (-1,0);
    \draw[gaugelink] (1,0) -- (1.45,0);
    \draw[gaugelink] (-1,0) arc (180:0:1);
    \draw[gaugelink] (1,0) arc (0:-180:1);
    \fill (-1,0) circle (2pt) (1,0) circle (2pt);
    \node at (0,-1.0) {gauge loop};
  \end{scope}
  \begin{scope}[xshift=13.5cm]
    \draw[gaugelink] (-1.45,0) -- (-1,0);
    \draw[gaugelink] (1,0) -- (1.45,0);
    \draw[ghostline] (-1,0) arc (180:0:1);
    \draw[ghostline] (1,0) arc (0:-180:1);
    \fill (-1,0) circle (2pt) (1,0) circle (2pt);
    \node at (0,-1.0) {ghost loop};
  \end{scope}
\end{tikzpicture}
\caption{Representative classes entering the gauge two-point tensor. The
Haar-measure contribution is an additional local quadratic insertion, not
shown as a loop. The continuum tensor is derived in Sec.~\ref{sec:g2pt}; the matter-only
lattice check passes, while the continuum-minus-lattice comparison and full
identity closure remain open.
}
\label{fig:gauge-pi-diagrams}
\end{figure*}
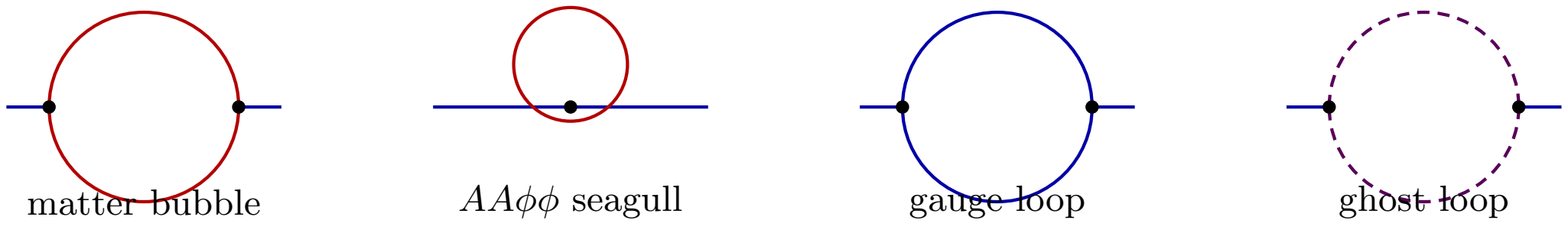

The one-loop gauge two-point diagrams fall into four topology classes: a
scalar-matter bubble, a scalar seagull, a pure-gauge loop, and a ghost loop
(Fig.~\ref{fig:gauge-pi-diagrams}). The Haar-measure term is a separate
local quadratic insertion, not a propagating loop.

\section{Continuum one-loop gauge two-point tensor}
\label{sec:g2pt}

This section supplies the continuum side of the gauge two-point comparison
for the symmetric background. The calculation includes the pure-gauge loop,
quartic-gauge tadpole, ghost loop, and scalar-doublet bubble and seagull,
with the fixed external-field convention and common infrared regulator used
in the lattice projections. Its algebraic and numerical checks are internal
to the implementation: 29 checks pass, but there has been no independent
code audit and no continuum-minus-lattice coefficient is formed. The
finite-regulator longitudinal term is therefore a diagnostic, not a claim
of completed Ward--Slavnov--Taylor closure.

\subsection{Conventions and tensor projections}

We use an $SU(2)$ gauge field coupled to one complex scalar doublet at
$\bar\rho=0$, with $m^2=0.4$, $g=g_0=0.65$, $C_A=2$, and
$\Tr(T^aT^b)=\tfrac12\delta^{ab}$. The external legs are the ordinary
quantum gauge field, not a background gauge field. Covariant Feynman gauge,
$\xi=1$, is fixed by $F^a=\partial_\mu A^a_\mu$ and its Faddeev--Popov
ghosts~\cite{FaddeevPopov1967,PascualTarrach1984,PeskinSchroeder1995}.
The common infrared regulator adds $\lambda_{\rm IR}^2$ to every massless
gauge and ghost denominator and leaves the massive scalar denominator
unchanged. The gauge-fixing quadratic term is kept separate from the
transverse Yang--Mills kernel
\begin{equation}
  K_{\mu\nu}(p)=p^2\delta_{\mu\nu}-p_\mu p_\nu,
\end{equation}
so the gauge-fixed tree inverse propagator is
$K_{\mu\nu}+p_\mu p_\nu/\xi$. We define the loop correction by
\begin{equation}
  \begin{aligned}
  \Gamma^{(2)ab}_{\mu\nu}(p)
    &=\delta^{ab}\left[K_{\mu\nu}(p)+\frac{p_\mu p_\nu}{\xi}
      +\Pi_{\mu\nu}(p)\right],\\
  \Pi_{\mu\nu}(p)
    &=A(p^2)\delta_{\mu\nu}
      +B(p^2)\frac{p_\mu p_\nu}{p^2}.
  \end{aligned}
  \label{eq:g2pt-decomposition}
\end{equation}

The tensor $\Pi$ is in the same normalization as the Wilson kernel and
the lattice Haar insertion
$\Delta\Pi^{\rm Haar}_{\mu\nu}=g_0^2/(6a^2)\,\delta_{\mu\nu}$.
In four dimensions the projectors used throughout this paper give
\begin{equation}
  \Pi_T=\frac{A}{p^2},\qquad
  \Pi_L=\frac{A+B}{p^2},\qquad
  p_\mu\Pi_{\mu\nu}=(A+B)p_\nu.
  \label{eq:g2pt-projectors}
\end{equation}
The loop expansion follows from
\begin{equation}
  \Gamma^{(1)}=\tfrac12\Tr\ln K_A-\Tr\ln M_{\rm gh}
  +\Tr\ln(-D^2+m^2),
  \label{eq:g2pt-effective-action}
\end{equation}
where $K_A$ is the gauge fluctuation operator and $M_{\rm gh}$ the ghost
operator. We use dimensional regularization in $d=4-2\epsilon$ and
$\MSbar$ subtraction at $\mu_R^2=0.4$~\cite{tHV,tHooft1973,BBDM}.
The loop measure is $\int_k=\mu_R^{2\epsilon}\int\dd^dk/(2\pi)^d$;
the common factor $1/(16\pi^2)$ is displayed explicitly in the scalar
integrals and restored in all projected numerical values.

\subsection{Loop integrands}

Let $D_1=k^2+M^2$, $D_2=(k+p)^2+M^2$, and define the three momenta at the
three-gauge vertex by $a=2k+p$, $b=p-k$, and $c=-(k+2p)$, so
$a+b+c=0$. Contracting two Euclidean three-gauge vertices gives
\begin{equation}
  N_{\mu\nu}=(d-2)a_\mu a_\nu+b_\mu c_\nu+c_\mu b_\nu
  +\delta_{\mu\nu}(b^2+c^2).
  \label{eq:g2pt-gluon-numerator}
\end{equation}
The four contributions per external color factor $\delta^{ab}$ are
\begin{align}
  \Pi^{\rm gl}_{\mu\nu}&=-\tfrac12g^2C_A\int_k
    \frac{N_{\mu\nu}}{D_1D_2},\nonumber\\
  \Pi^{\rm gh}_{\mu\nu}&=+g^2C_A\int_k
    \frac{(k+p)_\mu k_\nu}{D_1D_2},\nonumber\\
  \Pi^{\rm tad}_{\mu\nu}&=g^2C_A(d-1)\delta_{\mu\nu}
    I_1(\lambda_{\rm IR}),\nonumber\\
  \Pi^{\rm sc}_{\mu\nu}&=\tfrac{g^2}{2}\left[
    2\delta_{\mu\nu}I_1(m)\right.\nonumber\\
  &\left.\qquad-\int_k\frac{(2k+p)_\mu(2k+p)_\nu}{D_1D_2}\right].
  \label{eq:g2pt-loop-integrands}
\end{align}

Here $M^2=\lambda_{\rm IR}^2$ for gauge and ghost lines and $M^2=m^2$
for the scalar loop. The factor $1/2$ for the gluon loop, ghost-loop sign,
and color factors follow the standard gauge-fixed perturbative expansion
~\cite{Grozin2005,PascualTarrach1984,PeskinSchroeder1995}. In particular,
the relative weights follow from the quadratic and linear terms in the
operator logarithms, rather than from fitting the projected result.

\subsection{Tensor reduction and subtraction}

Define $T=\Pi_{\mu\mu}$ and
$P=p_\mu p_\nu\Pi_{\mu\nu}/p^2$. In $d$ dimensions,
\begin{equation}
  A=\frac{T-P}{d-1},\qquad B=P-A.
  \label{eq:g2pt-ab-reduction}
\end{equation}
The identities
\begin{equation}
  k^2=D_1-M^2,\qquad
  k\cdot p=\tfrac12(D_2-D_1-p^2)
  \label{eq:g2pt-denominator-identities}
\end{equation}
reduce the contractions to scalar tadpole and bubble masters; terms with
both $D_1$ and $D_2$ in the numerator are polynomial integrals and vanish
in dimensional regularization. The only required masters are
\begin{align}
  I_1(M)&=\int_k\frac{1}{D_1}
  =-\frac{M^2}{16\pi^2}\left[\frac1\epsilon+1-L\right],\nonumber\\
  B_0(p^2;M,M)&=\int_k\frac{1}{D_1D_2}
  =\frac{1}{16\pi^2}\left[\frac1\epsilon+b\right],\label{eq:g2pt-masters}
\end{align}
where
\begin{align}
  L&=\ln\frac{M^2}{\mu_R^2},\nonumber\\
  b&=-\int_0^1\dd x\,
    \ln\frac{M^2+x(1-x)p^2}{\mu_R^2},\nonumber\\
  &=2-L-\beta\ln\frac{\beta+1}{\beta-1},\nonumber\\
  \beta&=\sqrt{1+\frac{4M^2}{p^2}}.
  \label{eq:g2pt-bubble-finite}
\end{align}

For $M=0$, $b=2-\ln(p^2/\mu_R^2)$. This is the standard two-point
master-integral reduction~\cite{PassarinoVeltman1979}. The tensor
coefficients, which are rational in $d$, are expanded in $\epsilon$ before
subtracting the $1/\epsilon$ poles; this retains finite contributions from
$\epsilon\times1/\epsilon$ terms. In the expressions below, ``finite'' means
the remaining $\MSbar$ part after pole subtraction.

\subsection{Finite tensor coefficients and UV poles}

Writing $L_\lambda=\ln(\lambda_{\rm IR}^2/\mu_R^2)$,
$b_\lambda=b(p^2,\lambda_{\rm IR}^2)$,
$L_m=\ln(m^2/\mu_R^2)$, and $b_m=b(p^2,m^2)$, the finite tensor
coefficients are collected in Table~\ref{tab:g2pt-finite}.
\begin{table*}[!t]
\centering
\footnotesize
\setlength{\tabcolsep}{4pt}
\begin{tabular}{@{}l p{0.39\textwidth} p{0.39\textwidth}@{}}
\toprule
Contribution & $A_{\rm fin}$ & $(A+B)_{\rm fin}$\\
\midrule
Gluon loop & $\tfrac{C_Ag^2}{36}(-66L_\lambda\lambda_{\rm IR}^2+96\lambda_{\rm IR}^2b_\lambda+54\lambda_{\rm IR}^2-57b_\lambda p^2-2p^2)$ & $\tfrac{C_Ag^2}{4}(-14L_\lambda\lambda_{\rm IR}^2+4\lambda_{\rm IR}^2b_\lambda+6\lambda_{\rm IR}^2+b_\lambda p^2)$\\
Ghost loop & $\tfrac{C_Ag^2}{36}(6L_\lambda\lambda_{\rm IR}^2-12\lambda_{\rm IR}^2b_\lambda-18\lambda_{\rm IR}^2-3b_\lambda p^2-2p^2)$ & $\tfrac{C_Ag^2}{4}(2L_\lambda\lambda_{\rm IR}^2-2\lambda_{\rm IR}^2-b_\lambda p^2)$\\
Gauge tadpole & $C_Ag^2\lambda_{\rm IR}^2(3L_\lambda-1)$ & $C_Ag^2\lambda_{\rm IR}^2(3L_\lambda-1)$\\
Scalar doublet & $\tfrac{g^2}{18}(12L_mm^2+12m^2b_m+3b_mp^2+2p^2)$ & $0$\\
\bottomrule
\end{tabular}
\caption{Finite $\MSbar$ parts of the continuum one-loop tensor. Every entry carries an overall factor $1/(16\pi^2)$ and is per external color component. The coefficient of $1/\epsilon$ in each contribution to $A$, with the same overall factor, is given in the text.}
\label{tab:g2pt-finite}
\end{table*}
The coefficients of $1/\epsilon$, again with an overall $1/(16\pi^2)$,
are
\begin{align}
  A^{\rm gl}_{\rm pole}&=-\frac{C_Ag^2}{12}
    (19p^2-54\lambda_{\rm IR}^2),\\
  A^{\rm gh}_{\rm pole}&=-\frac{C_Ag^2}{12}
    (p^2+6\lambda_{\rm IR}^2),\\
  A^{\rm tad}_{\rm pole}&=-3C_Ag^2\lambda_{\rm IR}^2,\\
  A^{\rm sc}_{\rm pole}&=+\frac{g^2p^2}{6}.
  \label{eq:g2pt-poles}
\end{align}

The scalar-doublet contribution is exactly transverse. The three
gauge-sector longitudinal numerators combine to
\begin{equation}
  (A+B)_{\rm gauge}
  =C_Ag^2\lambda_{\rm IR}^2B_0(p^2;\lambda_{\rm IR},\lambda_{\rm IR}),
  \label{eq:g2pt-longitudinal}
\end{equation}
because the tadpole cancels all $I_1$ terms. Thus the regulator-induced
longitudinal term vanishes as $\lambda_{\rm IR}\to0$; at finite regulator
it contains a pole proportional to
$C_Ag^2\lambda_{\rm IR}^2/(16\pi^2\epsilon)$, corresponding to an
auxiliary gluon-mass counterterm.

\subsection{Internal validation}
\label{sec:g2pt-valid}

The implementation's validation suite records 29 checks, all passing.
The summary below distinguishes exact algebraic checks from numerical
cross-checks; these tests exercise the same implementation and do not
replace an independent audit.
\begin{table*}[!t]
\centering
\footnotesize
\begin{tabular}{@{}p{0.68\textwidth}p{0.23\textwidth}@{}}
\toprule
Check & Outcome\\
\midrule
Coefficient of $1/\epsilon$ in gauge-plus-ghost $\Pi_T$: $-\frac{5C_Ag^2}{3(16\pi^2)}$ & Exact\\\
Coefficient of $1/\epsilon$ in the scalar-doublet $\Pi_T$: $+\frac{T_Sg^2}{3(16\pi^2)}$, with $T_S=\tfrac12$ & Exact\\\
Massless gauge-plus-ghost finite part agrees with $-\frac{C_Ag^2}{16\pi^2}[-\tfrac53\ln(p^2/\mu_R^2)+\tfrac{31}{9}]$ & $<10^{-15}$ at $p^2=0.2,0.7,3$\\\
Scalar-doublet and massless gauge-plus-ghost tensors are transverse & Exact at $p^2=0.2,0.4,0.6$\\
Longitudinal sum equals Eq.~\eqref{eq:g2pt-longitudinal}, including its pole & $<10^{-16}$\\
Scalar finite part agrees with an independent Feynman-parameter integral & $10^{-16}$\\
Scalar $\Pi_T$ remains finite as $p^2\to0$ & No spurious $1/p^2$\\
Four-dimensional integrals of regulator-dependent second differences agree with the dimensional-regularization finite parts & $\lesssim10^{-10}$\\
\bottomrule
\end{tabular}
\caption{Summary of the 29 internal checks. The massless pole and finite $31/9$ comparison use standard one-loop QCD results~\cite{Grozin2005,PascualTarrach1984,GrossWilczek1973,Politzer1973}. Four-dimensional integral checks test the regulator-dependent terms from the same integrands, rather than providing an independent code implementation.}
\label{tab:g2pt-validation}
\end{table*}

\subsection{Benchmark projections and regulator dependence}

At $g_0=0.65$, $C_A=2$, $m^2=\mu_R^2=0.4$, the continuum projections at
three nonzero momenta and four regulator values are:
\begin{table*}[!t]
\centering
\footnotesize
\begin{tabular}{@{}cccccc@{}}
\toprule
$p^2$ & $\lambda_{\rm IR}/\mu_R$ & Gauge $\Pi_T$ & Scalar $\Pi_T$ & Total $\Pi_T$ & Total $\Pi_L$\\
\midrule
0.2 & $1/8$ & $-0.02216$ & $-0.00002$ & $-0.02218$ & $+4.05\times10^{-4}$\\
0.2 & $1/32$ & $-0.02439$ & $-0.00002$ & $-0.02441$ & $+2.79\times10^{-5}$\\
0.2 & $1/256$ & $-0.02461$ & $-0.00002$ & $-0.02463$ & $+4.40\times10^{-7}$\\
0.2 & $0$ & $-0.02461$ & $-0.00002$ & $-0.02463$ & $0$\\
0.4 & $1/8$ & $-0.01711$ & $-0.00004$ & $-0.01716$ & $+1.54\times10^{-4}$\\
0.4 & $1/32$ & $-0.01832$ & $-0.00004$ & $-0.01836$ & $+1.04\times10^{-5}$\\
0.4 & $1/256$ & $-0.01843$ & $-0.00004$ & $-0.01847$ & $+1.63\times10^{-7}$\\
0.4 & $0$ & $-0.01843$ & $-0.00004$ & $-0.01847$ & $0$\\
0.6 & $1/8$ & $-0.01391$ & $-0.00006$ & $-0.01397$ & $+8.25\times10^{-5}$\\
0.6 & $1/32$ & $-0.01474$ & $-0.00006$ & $-0.01480$ & $+5.52\times10^{-6}$\\
0.6 & $1/256$ & $-0.01481$ & $-0.00006$ & $-0.01487$ & $+8.68\times10^{-8}$\\
0.6 & $0$ & $-0.01482$ & $-0.00006$ & $-0.01488$ & $0$\\
\bottomrule
\end{tabular}
\caption{Continuum $\MSbar$ projections, dimensionless. The zero-regulator rows are evaluated directly from the massless expressions, not extrapolated.}
\label{tab:g2pt-benchmark}
\end{table*}
The zero-regulator transverse contributions at $p^2=0.4$ are listed in
Table~\ref{tab:g2pt-contributions}.
\begin{table}[t]
\centering
\footnotesize
\begin{tabular}{@{}lc@{}}
\toprule
Contribution & $\Pi_T$\\
\midrule
Gluon loop & $-0.01724$\\
Ghost loop & $-0.00119$\\
Gauge tadpole & $-0.00000$\\
Scalar doublet & $-0.00004$\\
Total & $-0.01847$\\
\bottomrule
\end{tabular}
\caption{Contributions to $\Pi_T$ at $p^2=0.4$ and zero regulator. The gluon loop dominates; the ghost and scalar contributions are about $7\%$ and $0.2\%$ of the total, respectively.}
\label{tab:g2pt-contributions}
\end{table}

\begin{figure*}[!t]
\centering
\begin{tikzpicture}
\begin{groupplot}[
  group style={group size=3 by 1,horizontal sep=1.15cm},
  width=0.285\textwidth,height=4.2cm,scale only axis,
  xmode=log,xmin=1.5e-5,xmax=2e-2,
  xtick={1e-5,1e-4,1e-3,1e-2},
  tick label style={font=\tiny},label style={font=\scriptsize},
  title style={font=\scriptsize},
  legend style={font=\tiny,fill=white,draw=gray,at={(0.03,0.97)},anchor=north west},
  xlabel={$x=(\lambda_{\rm IR}/\mu_R)^2$}
]
\nextgroupplot[title={Transverse; dotted: exact $\lambda_{\rm IR}=0$},
  ylabel={Total $\Pi_T$},ymin=-0.026,ymax=-0.012,
  ytick={-0.024,-0.020,-0.016,-0.012}]
\addplot[blue,thick,mark=*] coordinates {
 (1.52588e-5,-0.0246298) (6.10352e-5,-0.0246172) (2.44141e-4,-0.0245721)
 (9.76563e-4,-0.0244144) (3.90625e-3,-0.0238790) (1.5625e-2,-0.0221776)};
\addlegendentry{$p^2=0.2$}
\addplot[blue,densely dotted,forget plot] coordinates {(1.5e-5,-0.0246346) (2e-2,-0.0246346)};
\addplot[orange,thick,mark=square*] coordinates {
 (1.52588e-5,-0.0184706) (6.10352e-5,-0.0184641) (2.44141e-4,-0.0184409)
 (9.76563e-4,-0.0183591) (3.90625e-3,-0.0180778) (1.5625e-2,-0.0171553)};
\addlegendentry{$p^2=0.4$}
\addplot[orange,densely dotted,forget plot] coordinates {(1.5e-5,-0.0184730) (2e-2,-0.0184730)};
\addplot[green!60!black,thick,mark=triangle*] coordinates {
 (1.52588e-5,-0.0148743) (6.10352e-5,-0.0148699) (2.44141e-4,-0.0148541)
 (9.76563e-4,-0.0147985) (3.90625e-3,-0.0146061) (1.5625e-2,-0.0139665)};
\addlegendentry{$p^2=0.6$}
\addplot[green!60!black,densely dotted,forget plot] coordinates {(1.5e-5,-0.0148759) (2e-2,-0.0148759)};

\nextgroupplot[title={Regulator-induced longitudinal},ylabel={$|\Pi_L|$},
  ymode=log,ymin=1e-8,ymax=1e-3,ytick={1e-8,1e-6,1e-4}]
\addplot[blue,thick,mark=*] coordinates {
 (1.52588e-5,4.3968e-7) (6.10352e-5,1.7576e-6) (2.44141e-4,7.0147e-6)
 (9.76563e-4,2.7852e-5) (3.90625e-3,1.0879e-4) (1.5625e-2,4.0459e-4)};
\addlegendentry{$p^2=0.2$}
\addplot[orange,thick,mark=square*] coordinates {
 (1.52588e-5,1.6327e-7) (6.10352e-5,6.5277e-7) (2.44141e-4,2.6069e-6)
 (9.76563e-4,1.0370e-5) (3.90625e-3,4.0739e-5) (1.5625e-2,1.5389e-4)};
\addlegendentry{$p^2=0.4$}
\addplot[green!60!black,thick,mark=triangle*] coordinates {
 (1.52588e-5,8.6782e-8) (6.10352e-5,3.4699e-7) (2.44141e-4,1.3860e-6)
 (9.76563e-4,5.5171e-6) (3.90625e-3,2.1716e-5) (1.5625e-2,8.2466e-5)};
\addlegendentry{$p^2=0.6$}

\nextgroupplot[title={Extrapolation models at $p^2=0.4$},
  ylabel={$\Pi_T(x)-\Pi_T(0)$},ymin=0,ymax=0.0025,
  ytick={0,0.0005,0.0010,0.0015,0.0020,0.0025}]
\addplot[black,only marks,mark=*] coordinates {
 (1.52588e-5,2.4641e-6) (6.10352e-5,8.9499e-6) (2.44141e-4,3.2162e-5)
 (9.76563e-4,1.1395e-4) (3.90625e-3,3.9519e-4) (1.5625e-2,1.3177e-3)};
\addlegendentry{Exact continuum}
\addplot[blue,thick,domain=1.52588e-5:1.5625e-2,samples=80]
 {1.5705846e-5+0.1017528*x-1.1791481*x^2};
\addlegendentry{Quadratic, 3 regulators}
\addplot[orange,thick,domain=1.52588e-5:1.5625e-2,samples=80]
 {-2.04591e-8+0.0414219*x-0.1496765*x^2-0.0108808*x*ln(x)};
\addlegendentry{Quadratic $+x\ln x$, 6 regulators}
\end{groupplot}
\end{tikzpicture}
\caption{Continuum regulator checks. Left: total $\Pi_T$ versus
$x=(\lambda_{\rm IR}/\mu_R)^2$, with dotted lines showing the exact
zero-regulator limits. Center: the regulator-induced longitudinal
projection $|\Pi_L|$. Right: the exact continuum amplitude relative to its
zero-regulator value and two regulator-fit models at $p^2=0.4$.}
\label{fig:g2pt-regulator}
\end{figure*}
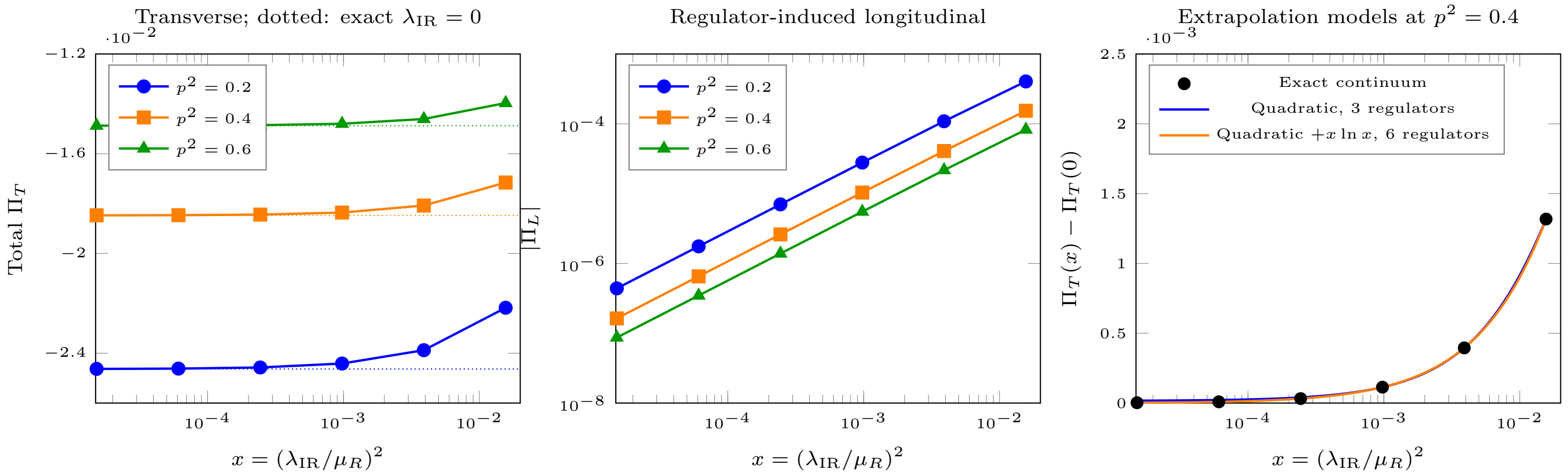

At $p^2=0.4$, the total $\Pi_T$ changes from $-0.017155$ at
$\lambda_{\rm IR}/\mu_R=1/8$ to $-0.018473$ at zero regulator, a $7.1\%$
shift, while $|\Pi_L|$ is at most $4.0\times10^{-4}$ across the sampled
grid. The exact function is not polynomial in $x$: its small-$x$
expansion contains an $x\ln x$ term whose coefficient in $\Pi_T$ is
$-0.0121$ at this momentum. Applying fit models to the exact analytic
continuum amplitude quantifies fit bias independently of QMC noise:
\begin{table*}[!t]
\centering
\footnotesize
\setlength{\tabcolsep}{4pt}
\begin{tabular}{@{}lccc@{}}
\toprule
Fit model (intercept minus exact) & $p^2=0.2$ & $p^2=0.4$ & $p^2=0.6$\\
\midrule
Quadratic, 3 regulators & $+3.19\times10^{-5}$ ($0.130\%$) & $+1.57\times10^{-5}$ ($0.085\%$) & $+1.04\times10^{-5}$ ($0.070\%$)\\
Quadratic, 4 regulators & $+1.99\times10^{-5}$ ($0.081\%$) & $+9.79\times10^{-6}$ ($0.053\%$) & $+6.50\times10^{-6}$ ($0.044\%$)\\
Quadratic, 5 regulators & $+1.34\times10^{-5}$ ($0.055\%$) & $+6.63\times10^{-6}$ ($0.036\%$) & $+4.40\times10^{-6}$ ($0.030\%$)\\
Quadratic, 6 regulators & $+9.81\times10^{-6}$ ($0.040\%$) & $+4.84\times10^{-6}$ ($0.026\%$) & $+3.22\times10^{-6}$ ($0.022\%$)\\
Linear plus $x\ln x$, 4 regulators & $-3.94\times10^{-6}$ ($0.016\%$) & $-1.24\times10^{-6}$ ($0.007\%$) & $-6.20\times10^{-7}$ ($0.004\%$)\\
Quadratic plus $x\ln x$, 6 regulators & $-8.34\times10^{-8}$ ($0.000\%$) & $-2.05\times10^{-8}$ ($0.000\%$) & $-9.00\times10^{-9}$ ($0.000\%$)\\
\bottomrule
\end{tabular}
\caption{Intrinsic intercept bias of each regulator-fit model applied to
the exact continuum $\Pi_T$; percentages are relative to the exact value.}
\label{tab:g2pt-fit-bias}
\end{table*}
The three-regulator quadratic baseline has a same-sign bias of order
$0.1\%$ at all three momenta, decreasing as the fit window expands. Fits
that include $x\ln x$ reduce this analytic bias to the $10^{-6}$--$10^{-8}$
range. This does not establish the best fit model for lattice data, but it
shows why regulator-model sensitivity must be reported separately from
integration errors.

At $p^2=0.4$ and $\lambda_{\rm IR}/\mu_R=1/32$, the continuum longitudinal
projection is $\Pi_L=1.04\times10^{-5}$, or
$p_\mu\Pi_{\mu\nu}=4.1\times10^{-6}p_\nu$; for
$p\propto(1,2,3,4)/\sqrt{30}$, the largest component is approximately
$1.9\times10^{-6}$. The raw lattice follow-up reports a projected Ward
component $-0.01198\pm0.00052$ at that setting. If both quantities use the
same normalization, the continuum regulator effect is roughly three orders
of magnitude smaller. Because agreement of conventions and normalizations
has not been checked, this comparison is conditional: it is not evidence
that the lattice identity fails, nor does it assign the lattice residual to
a particular omitted or discretization contribution.

The continuum $\Pi_T(\lambda_{\rm IR}\to0)$ values are now available as the
continuum side of gauge-field normalization matching. The matched difference
$\Pi_{\overline{\rm MS},\rm ren}-\Pi_{\rm lat}$, independent audit,
gauge-coupling matching, and Ward--Slavnov--Taylor closure remain open; no
gauge-field coefficient is entered into the coefficient template.

\section{Discussion: scope, limitations, and completion gates}
\label{sec:discussion}

\subsection{What the present results establish}
\label{subsec:established}

The principal advance is the construction and audit of a usable algebraic
framework for a lattice $\SUtwo$--Higgs calculation. The nearest-neighbor
scalar-link action and Wilson plaquette action define finite-spacing
vertices without ambiguity about link orientation, Fourier phases, or
generator normalization. Their derivatives agree with independent local
expansions to the precision listed in Table~\ref{tab:vertex-audit}. The
operator basis and fixed kinematic routes also permit sector-by-sector
matching without conflating a local contact with momentum-dependent terms.

Within the scalar-only scope, the radial potential matching through
$\rho^{12}$ is numerically reproduced and its quadrature convergence has
been checked. The Higgs slope projection is likewise a finite result for
its specified two-point sector. The selected gauge-linked projections show
that the fixed-route integrators can return finite estimates and that
paired QMC is practical for suppressing variance in continuum-minus-lattice
differences. The continuum one-loop gauge two-point tensor at the symmetric
background is now available and passes 29 internal checks; it supplies the
continuum side only, since an independent audit and the lattice comparison
remain outstanding. These are necessary technical results for a larger
calculation.

They are not sufficient to specify a fully matched gauged effective action.
The scalar-only coefficient table omits gauge and ghost loops. The value of
$\delta Z_H$ does not determine the gauge-field wave-function coefficient
or the coupling renormalization. Higher-point pilot intercepts do not
become coefficients merely by extrapolating their IR regulator to zero;
the full counterterm normalization, other diagrams, UV subtractions, and
identity relations must also be satisfied.

\subsection{Why the gauge potential cannot be inferred by a scalar Taylor series}
\label{subsec:nonanalyticity}

For a constant radial background in background Feynman gauge, the one-loop
lattice determinant takes the schematic form
\begin{align}
 V_{1,\rm lat}(\rho)=\frac12\int_{\BZ}\frac{\dd^4k}{(2\pi)^4}
 \bigl[&\log(\hat k^2+M_L^2)
 +3\log(\hat k^2+M_T^2+M_A^2)\nonumber\\
 &+6\log(\hat k^2+M_A^2)\bigr],
 \label{eq:constant-det-lat}
\end{align}
up to a background-independent constant. The continuum expression must retain
the $d$-dependent vector/ghost multiplicity through dimensional
regularization and $\MSbar$ subtraction; setting it to its four-dimensional
value before subtracting can discard a finite term
\cite{Abbott,tHV,BBDM}. In four dimensions the net gauge multiplicity
corresponds to two transverse polarizations for each of the three gauge
bosons after ghost cancellation.

The gauge contribution includes the familiar nonanalytic behavior
$M_A^4\log M_A^2\propto\rho^4\log\rho^2$ near $\rho=0$
\cite{CW}. Thus a Taylor expansion of the constant-background determinant
at the symmetric point is not a valid way to infer the full set of local
gauge-linked coefficients. The amplitude projectors must be applied to
properly renormalized, nonexceptional Green functions, and nonlocal or
nonanalytic pieces must be kept distinct from local matching terms.

\subsection{Open gates}
\label{subsec:open-gates}

The remaining work is organized as a sequence of physics and numerical
gates rather than as a single additional integration run:
\begin{enumerate}
\item \textbf{Close the gauge two-point sector.} Compare the continuum
tensor of Sec.~\ref{sec:g2pt} with the complete lattice matter, gauge,
ghost, seagull, and Haar contributions in the same conventions. Audit the
continuum calculation independently, form the matched difference, and
extract transverse normalization before inferring $\delta Z_A$.

\item \textbf{Establish the Ward--Slavnov--Taylor closure.} Check the
  longitudinal and transverse identities with a regulator prescription
  whose finite-regulator effects are quantified and whose zero-regulator
  limit is controlled.
\item \textbf{Complete coefficient-level projections.} Re-evaluate the
  gauge-linked four-, six-, and higher-point sectors with enough independent
  scrambles, held-out regulator values, and alternative fit bases to
  estimate model sensitivity. Include all diagrams and UV subtractions
  required by the chosen truncation.
\item \textbf{Match the gauge coupling and field normalization.} The
  benchmark $g_0$ is a fixed input, not a determination of $g_0(a)$.
  Coupling and gauge-field matching must be consistent with the scalar
  coefficient basis.
\item \textbf{Populate and independently validate the coefficient set.}
  The full coefficient template should remain null until every required
  sector and identity gate passes; reproducibility checks should regenerate
  the published values from archived inputs and outputs.
\item \textbf{Return to the decay observable.} Only after the preceding
  matching work can the gauge--Goldstone--ghost fluctuation determinant,
  collective-coordinate normalization, and continuum and volume limits be
  combined with the Part I scalar saddle for a physical rate.
\end{enumerate}
This ordering avoids treating a precise partial coefficient as evidence of
a precise decay prefactor. It also separates coefficient matching from the
additional numerical tasks involved in a bounce calculation and a
renormalized determinant.

\subsection{Illustrative computational budget}
An illustrative cost can be estimated for a representative run with eight
scrambles, $2^{12}$ points per scramble, six regulator values, and three
routes. This gives
\begin{equation}
  N_{\rm pt}=8\times2^{12}\times6\times3=589{,}824
\end{equation}
sample points per regularization. Counting one continuum and one lattice
evaluation at each point gives $2N_{\rm pt}=1{,}179{,}648$ loop-point
evaluations. If $t_L$ is the serial time per loop-point evaluation on one
side, the estimated CPU cost is
\begin{equation}
  H_{\rm CPU}(L)=\frac{2N_{\rm pt}t_L}{3600}.
\end{equation}
Using the quoted per-point timings gives the following estimates:
\begin{center}
\small
\begin{tabular}{@{}rccc@{}}
\toprule
 $L$ & $t_L$ (s/pt) & CPU-h & One-core days\\
\midrule
8  & 0.12 & 39.3 & 1.64\\
10 & 0.95 & 311.3 & 12.97\\
12 & 6.35 & 2080.8 & 86.70\\
\bottomrule
\end{tabular}
\\[2pt]
\footnotesize\emph{Estimated serial CPU cost for the stated sample count,
counting one continuum and one lattice evaluation per sample point.}
\end{center}
Thus, one run of this size at $L=12$ costs about $2.08\times10^3$ CPU-hours,
or 86.7 days on one continuously occupied core. These are extrapolations
from a single timing point at each $L$, not end-to-end measurements; hardware
details and timing variability should be recorded before treating them as a
portable forecast. Parallel wall time cannot be inferred by dividing by the
number of cores without measured scaling. The estimate covers only this
one run design, not the full remaining program: additional scrambles,
regulator and fit-window checks, other operator sectors, independent audits,
Ward--Slavnov--Taylor closure, spacing refinement, and the fluctuation
determinant are excluded. In particular, the exploratory $L=10$ and $L=12$
higher-point values are cancellation dominated and have relative errors of
roughly 10--17\% at the current low statistics; the cost estimate does not
establish that they are stable or sufficiently precise. One thousand CPU single P-core days may be required for completion.

\subsection{Relation to Part I}
\label{subsec:part-one-relation}

Part I remains the benchmark for the global-$O(4)$ scalar solver and for
the interpretation of its continuum-subtracted determinant. Part II does
not revise the scalar bounce action, the low-mode spectrum, or the
prescription-level radial determinant reported there. Instead, it documents
the perturbative matching layer needed to connect a finite lattice action
to continuum-renormalized parameters and to extend the Hessian analysis to
gauge, Goldstone, and ghost fluctuations. The distinction is essential:
matching a set of local vertices is not the same calculation as evaluating
the complete gauge-sector determinant around the inhomogeneous bounce.

The two parts therefore form a sequence of complementary checks. The scalar
benchmark tests the saddle and scalar fluctuation machinery; the present
paper documents the action and selected perturbative projections.
Neither result should be advertised as a measured electroweak vacuum-decay
rate, and the scalar-only values in Table~\ref{tab:scalar-matching} should
not be combined with the Part I determinant as though they formed a full
matched theory.

\section{Conclusions}
\label{sec:conclusion}

We have presented the lattice-action and matching framework developed as
the next stage of the four-dimensional Higgs fluctuation solver. The
Wilson-plaquette and scalar-link conventions fix the finite-spacing
vertices, while exact derivatives of the link, plaquette, potential, and
gauge-fixing terms provide a validated algebraic basis. Fixed routes and
projectors define how continuum-renormalized and lattice amplitudes are to
be compared.

For the stated scalar-only scope, the radial potential coefficients through
$\rho^{12}$ are complete at the benchmark, and the Higgs kinetic slope has
been evaluated. Gauge-linked contact projections provide useful pilots,
including an eight-point paired run, but their reported errors do not
include fit-model uncertainty and they do not establish a complete local
coefficient set. The continuum one-loop gauge two-point tensor at the
symmetric background is now derived and internally validated, but it has not
been independently audited or compared with the lattice. Full
Ward--Slavnov--Taylor closure, gauge-coupling matching, and the complete
gauge-sector determinant remain open. The full coefficient template
therefore stays unpopulated, and no fully matched $\SUtwo$ decay prefactor
is reported.

The practical conclusion is that Part II supplies the infrastructure and
the diagnostic record required to proceed, while keeping the physical claim
at the level supported by the calculations. The next milestone is an
independent audit and convention-matched lattice comparison for the gauge
two-point tensor, followed by Ward--Slavnov--Taylor closure and the remaining
coefficient and determinant work.

\appendix

\section{Notation and symbols}
\label{app:notation}

This appendix collects the variables, indices, operators, and abbreviations
used throughout the manuscript. A few letters are reused in different
contexts; those meanings are distinguished explicitly below.

\subsection*{Indices, fields, and model parameters}
\begin{description}
\item[$x,y$] Lattice sites; $x+a\hat\mu$ is the nearest neighbor of $x$ in
  direction $\mu$.
\item[$a$] Lattice spacing when unindexed. In $A^a_{x\mu}$ or
  $\delta^{ab}$, $a,b$ denote adjoint $SU(2)$ indices; $a_\mu,b_\mu,c_\mu$ in
  the continuum gauge-loop calculation are momentum vectors.
\item[$\mu,\nu$] Euclidean spacetime-direction indices, ranging from 1 to 4
  in the lattice theory; $\hat\mu$ is the unit vector in direction $\mu$.
  The symbol $\mu_R$ is instead the renormalization scale.
\item[$L$] Linear lattice extent in captions such as $L=20$; in the
  continuum master-integral formulas, $L$ instead denotes
  $\ln(M^2/\mu_R^2)$.
\item[$A,B$] Real-scalar component indices, ranging from 1 to 4. In the
  continuum gauge-tensor decomposition, $A(p^2)$ and $B(p^2)$ instead name
  scalar form factors; $A^a_{x\mu}$ denotes the gauge field.
\item[$H_x,\phi_x^A,\rho_x$] The complex scalar doublet, its four real
  components, and its radial magnitude, respectively, with
  $\rho_x^2=2H_x^\dagger H_x=\sum_A(\phi_x^A)^2$.
\item[$\bar\rho_x,h,\chi^a$] Radial background, radial Higgs fluctuation,
  and three transverse Goldstone fluctuations; in a radial background
  $\phi^4=\bar\rho+h$ and $\phi^a=\chi^a$ for $a=1,2,3$.
\item[$O(4),\SUtwo,R_\xi$] Global four-component orthogonal symmetry,
  special-unitary gauge group with two colors, and background gauge fixing
  with gauge parameter $\xi$.
\item[$\MSbar,\BZ,\dd,\Tr$] Modified-minimal-subtraction scheme,
  lattice Brillouin zone, integration differential, and matrix trace,
  respectively; $\operatorname{ReTr}$ denotes the real part of a trace.
  QMC means quasi--Monte Carlo and SE means standard error.
\item[$\mu_{\rm sq},m_0^2,m^2,\mu_R^2$] The quadratic tree-potential
  parameter, false-vacuum mass squared $m_0^2=-\mu_{\rm sq}$, the scalar
  mass squared in continuum loop formulas (equal to $m_0^2$ here), and the
  squared $\MSbar$ subtraction scale. At the benchmark
  $m_0^2=m^2=\mu_R^2=0.4$; $m_0$ is the positive square root of $m_0^2$.
\item[$\lambda,\gamma,\delta$] Tree-potential couplings multiplying
  $\rho^4/4$, $\rho^6/6$, and $\rho^8/8$, respectively. Here $\delta$ is
  a coupling, not a Kronecker delta.
\item[$\delta m^2,\delta\lambda,\delta\gamma,\delta\delta,
  \delta c_{10},\delta c_{12}$] Counterterm coefficients for the local
  operators $\rho^2/2,\rho^4/4,\rho^6/6,\rho^8/8,\rho^{10}/10,$ and
  $\rho^{12}/12$, respectively. The prefixed $\delta$ denotes a
  counterterm; in particular, $\delta\delta$ is the counterterm for the
  octic coupling $\delta$.
\item[$\delta Z_H$] Higgs wave-function (kinetic) counterterm coefficient.
\item[$V_0,V_{\rm ct}$] Tree-level scalar potential and its local
  counterterm potential.
\item[$V_{1,\rm lat}^{\rm scalar},V_{1,\rm lat}$] Scalar-only and
  gauge-background one-loop lattice effective potentials, respectively.
\item[$M_L^2(\rho),M_T^2(\rho),M_A^2$] Radial scalar curvature,
  transverse/Goldstone curvature, and gauge-boson mass parameter; the last
  is $q^2\rho^2$.
\item[$P_L(z),P_T(z),z$] Radial and transverse scalar mass shifts and the
  expansion variable $z=\rho^2$ used in the scalar-only effective
  potential. These $P_{L,T}$ are not the gauge-tensor projectors.
\end{description}

\subsection*{Lattice and gauge-theory objects}
\begin{description}
\item[$T^a,\sigma^a$] Fundamental $SU(2)$ generators and Pauli matrices,
  with $T^a=\sigma^a/2$ and $\Tr(T^aT^b)=\delta^{ab}/2$.
\item[$g_0,g,q$] Bare lattice gauge coupling, continuum gauge coupling,
  and the shorthand $q=g_0/2$. The continuum two-point calculation sets
  $g=g_0$ at the stated benchmark; this does not constitute coupling
  matching.
\item[$U_{x\mu},U_{x,\mu\nu}$] Oriented link matrix and ordered plaquette
  matrix. $H_+$ in the link diagram abbreviates $H_{x+a\hat\mu}$.
\item[$S_g,S_{H,\rm link},S_{H,x\mu},S$] Wilson gauge action, scalar
  nearest-neighbor link action, its contribution from link $(x,\mu)$, and
  the full perturbative action.
\item[$S_{\rm gf},S_{\rm gh},F_x^a$] Gauge-fixing action, Faddeev--Popov
  ghost action, and gauge-fixing function. $D$ in $F_x^a=(DA)_x^a-\xi
  q\bar\rho_x\chi_x^a$ is the lattice backward divergence; in the
  scalar-only loop potential, $D$ instead abbreviates $\hat k^2$.
\item[$c_x^a,\bar c_x^a,\omega_y^b,M_{xy}^{ab}$] Ghost, antighost,
  infinitesimal gauge-transformation parameter, and Faddeev--Popov operator
  $M_{xy}^{ab}=\delta F_x^a/\delta\omega_y^b$.
\item[$\xi$] Gauge-fixing parameter; $\xi=1$ is Feynman gauge.
\item[$J(A)$] Haar-measure density for logarithmic link coordinates. The
  associated local quadratic insertion is
  $\Delta\Pi_{\mu\nu}^{\rm Haar}$; $\delta_{\mu\nu}$ and $\delta^{ab}$
  denote Kronecker deltas.
\item[$F_{\mu\nu}^a,D_\mu$] Continuum non-Abelian field strength and
  continuum covariant derivative; $D^2$ is its Laplacian in the operator
  logarithm. The unindexed $D$ has the context-specific meanings given above.
\item[$r$] Number of gauge-field derivatives (and hence gauge legs) in the
  scalar-link vertex generator.
\item[$\hat k_\mu,\hat p_\mu$] Lattice momenta,
  $\hat k_\mu=(2/a)\sin(ak_\mu/2)$ and analogously for $p$;
  $\hat k^2=\sum_\mu\hat k_\mu^2$. $\BZ$ denotes the lattice
  Brillouin zone.
\item[$\theta_\mu$] Link-momentum phase in the kinetic-counterterm
  vertices, $\theta_\mu=ap_\mu/2$.
\item[$\bar\rho,\rho$] In the local-vertex and continuum gauge checks,
  $\bar\rho$ is the constant background; $\rho$ without a bar is the
  radial field or its magnitude.
\end{description}

\subsection*{Matching, momenta, and numerical projections}
\begin{description}
\item[$\Gamma^{(n)}$] One-particle-irreducible (1PI) $n$-point vertex;
  $\Delta\Gamma^{(n)}=\Gamma^{(n)}_{\MSbar,\rm ren}-\Gamma^{(n)}_{\rm lat}$
  is the continuum-renormalized-minus-lattice difference at common external
  kinematics. $\Gamma^{(1)}$ in the continuum gauge subsection instead
  denotes the one-loop effective action.
\item[$p,k$] Continuum external and loop momenta; $p^2=\sum_\mu p_\mu^2$.
  The symbol $s$ labels the selected external invariant $p^2$, while
  $\hat n$ is the fixed unit routing direction in $p=\sqrt{s}\,\hat n$.
\item[$p_{H^\dagger},p_H,p_A$] The three incoming momenta in the
  asymmetric $H^\dagger H A$ routing; they sum to zero.
\item[$\Sigma^{(2)}_{\MSbar,\rm ren},\Sigma^{(2)}_{\rm lat}$] Continuum
  renormalized and lattice Higgs two-point self-energies used to project
  $\delta Z_H$.
\item[$\mathcal{A}_{2n},\,\bar{s},\,\delta c_{2n}$] Contracted $2n$-scalar
  amplitude, scaled kinematic variable used for its local-momentum fit, and
  the coefficient of $\rho^{2n}/(2n)$. Here $n$ counts scalar pairs.
\item[$d_0,d_1,d_2$] Fit coefficients in
  $\mathcal{A}_{2n}(\bar s)=d_0+d_1\bar s+d_2\bar s^2$; $d_0$ is the
  extrapolated local intercept.
\item[$c_0,c_1,c_2$] Coefficients of a quadratic fit to a two-point
  function in its continuum or lattice kernel coordinate.
\item[$T_{\mu\nu},L_{\mu\nu},P_T^{\mu\nu},P_L^{\mu\nu}$] Transverse
  and longitudinal momentum tensors and their four-dimensional projection
  tensors in the matching-method section. There $K=p$ in the continuum and
  $K=\hat p$ on the lattice.
\item[$\lambda_{\rm IR},x$] Auxiliary mass used to regulate massless
  gauge and ghost lines, and the regulator-fit variable
  $x=(\lambda_{\rm IR}/\mu_R)^2$. Elsewhere $x$ denotes a lattice site or
  the Feynman parameter in a bubble integral, as specified by context.
\item[$\epsilon,d$] Dimensional-regularization parameter and spacetime
  dimension, related by $d=4-2\epsilon$.
\item[$I_n^{\rm lat},I_n^{\MSbar},\Delta v_n$] Lattice and subtracted
  continuum scalar loop moments and the coefficient of $z^n$ in the
  one-loop potential difference. In the Bessel representation, $t$ is a
  Schwinger parameter, $I_0$ is the modified Bessel function, and $j,n$ are
  integer expansion or moment indices. The unindexed $I$ in the schematic
  Higgs diagram denotes a tadpole loop integral.
\item[$\Gamma(n)$] Euler gamma function when its argument is an integer
  moment index, not a 1PI vertex.
\end{description}

\subsection*{Continuum gauge two-point calculation}
\begin{description}
\item[$\Gamma^{(2)ab}_{\mu\nu},K_{\mu\nu}(p),\Pi_{\mu\nu}$] Gauge
  inverse two-point function, transverse Yang--Mills tree kernel
  $K_{\mu\nu}=p^2\delta_{\mu\nu}-p_\mu p_\nu$, and its loop correction.
  The gauge-fixing tree term $p_\mu p_\nu/\xi$ is written separately.
\item[$\Pi_T,\Pi_L,A(p^2),B(p^2)$] Transverse and longitudinal scalar
  projections of the loop tensor and the form factors in
  $\Pi_{\mu\nu}=A\delta_{\mu\nu}+B p_\mu p_\nu/p^2$; in four dimensions
  $\Pi_T=A/p^2$ and $\Pi_L=(A+B)/p^2$.
\item[$K_A,M_{\rm gh},D^2$] Gauge fluctuation operator, ghost operator,
  and covariant Laplacian appearing in the one-loop operator logarithms.
\item[$D_1,D_2,M$] Propagator denominators
  $D_1=k^2+M^2$ and $D_2=(k+p)^2+M^2$, and the internal-line mass
  parameter. $M^2=\lambda_{\rm IR}^2$ for gauge/ghost lines and $M^2=m^2$
  for scalar lines.
\item[$a_\mu,b_\mu,c_\mu,N_{\mu\nu}$] Three momenta at the three-gauge
  vertex, $a=2k+p$, $b=p-k$, $c=-(k+2p)$, and the numerator from contracting
  the two vertices. These momentum-vector $a,b,c$ are unrelated to the
  lattice spacing and to scalar fit coefficients.
\item[$\Pi^{\rm gl},\Pi^{\rm gh},\Pi^{\rm tad},\Pi^{\rm sc}$] Gluon
  loop, ghost loop, gauge tadpole, and scalar-doublet contributions to the
  one-loop tensor.
\item[$T,P$] In the tensor-reduction subsection only,
  $T=\Pi_{\mu\mu}$ and $P=p_\mu p_\nu\Pi_{\mu\nu}/p^2$; these are
  contractions, not the matching tensors $T_{\mu\nu}$ and $P_T^{\mu\nu}$.
\item[$I_1(M),B_0(p^2;M,M)$] Tadpole and scalar bubble master integrals.
  $L=\ln(M^2/\mu_R^2)$, $b$ is the finite part of the bubble, and
  $\beta=\sqrt{1+4M^2/p^2}$ is the parameter in its closed form.
\item[$L_\lambda,b_\lambda,L_m,b_m$] The logarithm and bubble finite part
  evaluated for regulator mass $\lambda_{\rm IR}$, and the corresponding
  quantities evaluated for scalar mass $m$.
\item[$C_A,T_S$] Adjoint Casimir and scalar-representation index; for the
  conventions here $C_A=2$ and $T_S=1/2$.
\item[$\int_k$] Dimensionally regulated loop measure
  $\mu_R^{2\epsilon}\int\dd^dk/(2\pi)^d$; an overall $1/(16\pi^2)$ is
  displayed in the scalar master integrals.
\end{description}

\begin{figure*}[!t]
\section{Diagram conventions and free propagators}
\label{app:diagram-conventions}

Figure~\ref{fig:appendix-line-conventions} and
Table~\ref{tab:appendix-free-propagators} summarize the line/vertex key and
free propagators used in the calculations.\vspace{\baselineskip}

\centering
\begin{tikzpicture}[x=1cm,y=1cm]
  \draw[higgs] (0,0) -- (2,0); \node[right] at (2.2,0) {radial Higgs $h=\phi^4-\bar\rho$ (``$L$'' channel)};
  \draw[gold] (0,-0.8) -- (2,-0.8); \node[right] at (2.2,-0.8) {Goldstone $\chi^a$ (``$T$'' channel); scalar-only loops: all 4 real components};
  \draw[gauge] (0,-1.6) -- (2,-1.6); \node[right] at (2.2,-1.6) {lattice gauge field $A^a_{x\mu}$ (link midpoint), $U=e^{-iag_0A^aT^a}$};
  \draw[scal] (0,-2.4) -- (2,-2.4); \node[right] at (2.2,-2.4) {fundamental doublet $H_x$ (arrow: $H\to H^\dagger$ flow)};
  \draw[ghost] (0,-3.2) -- (2,-3.2); \node[right] at (2.2,-3.2) {Faddeev--Popov ghost $c^a,\bar c^a$};
  \fill (0.9,-4.0) circle (2pt); \node[right] at (2.2,-4.0) {local vertex from the proposed action (exact lattice vertex generator)};
  \ctv{0.9,-4.8} \node[right] at (2.2,-4.8) {matching counterterm vertex ($\delta Z_H,\delta m^2,\delta\lambda,\delta\gamma,\delta\delta,\delta c_{10},\delta c_{12}$)};
  \node[font=\scriptsize] at (0.9,-5.6) {$V,P,U$}; \node[right] at (2.2,-5.6) {insertion labels: $V=hA\chi$ mixing, $P=hhAA$ seagull, $U_n$ Goldstone-Hessian vertices};
\end{tikzpicture}
\caption{Line and vertex conventions used throughout this manuscript. All
external momenta are incoming; site fields use $e^{+ip\cdot x}$ and gauge
fields are placed at link midpoints. Lattice kernels use
$\hat k_\mu=(2/a)\sin(ak_\mu/2)$.}
\label{fig:appendix-line-conventions}
\end{figure*}

\begin{table*}[!t]
\centering
\small
\begin{tabular}{@{}lll@{}}
\toprule
Free propagator & Expression & Notes\\
\midrule
Higgs/Goldstone & $\delta^{AB}/(\hat k^2+m_0^2)$ & $m_0^2=-\mu_{\rm sq}$; benchmark $a=0.8$, $m_0^2=0.4$\\
Gauge ($\xi=1$) & $\delta_{\mu\nu}/(\hat k^2+\lambda_{\rm IR}^2)$ & common IR shift on both regularizations\\
Ghost & $1/(\hat k^2+\lambda_{\rm IR}^2)$ & same shift\\
Ghost at $\bar\rho\neq0$ & $\delta^{ab}/(\hat k^2+\xi q^2\bar\rho^2)$ & $q=g_0/2$\\
\bottomrule
\end{tabular}
\caption{Free propagators associated with the line conventions in
Fig.~\ref{fig:appendix-line-conventions} and used in the one-loop
calculations.}
\label{tab:appendix-free-propagators}
\end{table*}

\begin{table*}[!t]
\section{Status dashboard}
\label{app:status-dashboard}

Table~\ref{tab:status} condenses the evidence. ``Complete'' always refers to
the specific scope in the table, not to full gauge-theory matching.
\centering
\scriptsize
\begin{tabular}{@{}p{0.20\textwidth}p{0.18\textwidth}p{0.24\textwidth}p{0.29\textwidth}@{}}
\toprule
Sector & Status & Reported result & Limitation\\
\midrule
Lattice action and exact vertex generators & Audited formulation & Largest listed discrepancy $2.4\times10^{-6}$ & Not yet consumed by a complete matching integrator or production determinant\\
Routings and local projectors & Algebraically audited & Tree-slope projector error $1.1\times10^{-15}$ & Does not test a one-loop Ward identity\\
Scalar-only radial coefficients & Complete for selected scope & Six coefficients through $\rho^{12}$; Table~\ref{tab:scalar-matching} & Gauge, link, and ghost loops excluded\\
Higgs kinetic coefficient & Evaluated two-point sector & $\delta Z_H=0.0329117974\pm3.6\times10^{-7}$ & QMC-only error; three-point fit has no independent extra-point check\\
Gauge-linked four-Higgs projection & Pilot & $0.2540212\pm0.0007361$ & Held-out residuals have same sign; no model-systematic estimate\\
Gauge-linked six-Higgs projection & Pilot & Earlier $-2.2646\pm0.0702$; rerun $-2.2199\pm0.1526$ & Four-scramble runs are underpowered for stability gates\\
Gauge-linked eight-Higgs projection & Paired run complete & $-19.8545\pm1.3014$ & Five-point window shift is about $2.4$ paired SE\\
Gauge-linked ten- and twelve-Higgs projections & Exploratory & $O(10^4)$ and $O(10^6)$ central-value magnitudes & Four points per scramble; not interpreted as coefficients\\
Gauge two-point continuum tensor & Internally validated & Sec.~\ref{sec:g2pt}; $\Pi_T(0.4,0)=-0.01847$ & Independent audit, lattice subtraction, and identity closure remain open\\

Full coefficient template & Unpopulated & Full-matching flag remains false & Full $\SUtwo$ matching and determinant are not established\\
\bottomrule
\end{tabular}
\caption{Status of the lattice-to-continuum matching program. Quoted errors
for gauge-linked amplitudes are integration errors only; the continuum
gauge two-point checks are internal validation, not an independent audit.}

\label{tab:status}
\end{table*}
\clearpage

\begin{acknowledgments}
We acknowledge and thank OpenAI Prism, Anthropic, Google Cloud Shell, and Microsoft Azure
Ubuntu for their assistance with coding, running simulations, and manuscript
preparation.
\end{acknowledgments}


\begin{thebibliography}{99}

\bibitem{Coleman1977}
S. Coleman, ``The Fate of the False Vacuum: Semiclassical Theory,''
Phys. Rev. D \textbf{15}, 2929 (1977).

\bibitem{CallanColeman1977}
C. G. Callan, Jr. and S. Coleman, ``The Fate of the False Vacuum. II. First
Quantum Corrections,'' Phys. Rev. D \textbf{16}, 1762 (1977).

\bibitem{CookPartI}
M. S. Cook, ``Higgs Fluctuation Solver in Four Dimensions: Part I---A
Matrix-Free Global-Scalar Bounce and Determinant-Renormalization Audit,''
arxiv.org 2610.30341 (2026).

\bibitem{Wilson1974}
K. G. Wilson, ``Confinement of Quarks,'' Phys. Rev. D \textbf{10}, 2445
(1974).

\bibitem{LW}
M. L\"uscher and P. Weisz, ``Efficient numerical techniques for perturbative
lattice gauge theory computations,'' Nucl. Phys. B \textbf{266}, 309--356
(1986).

\bibitem{Abbott}
L. F. Abbott, Nucl. Phys. B \textbf{185}, 189 (1981).

\bibitem{Reisz}
T. Reisz, ``A power counting theorem for Feynman integrals on the lattice,''
Commun. Math. Phys. \textbf{116}, 81--126 (1988).

\bibitem{Kawai}
H. Kawai, R. Nakayama, and K. Seo, ``Comparison of the lattice $\Lambda$
parameter with the continuum $\Lambda$ parameter in massless QCD,''
Nucl. Phys. B \textbf{189}, 40--62 (1981).

\bibitem{Taylor}
J. C. Taylor, Nucl. Phys. B \textbf{33}, 436 (1971).

\bibitem{FaddeevPopov1967}
L. D. Faddeev and V. N. Popov, ``Feynman diagrams for the Yang--Mills
field,'' Phys. Lett. B \textbf{25}, 29--30 (1967).

\bibitem{PassarinoVeltman1979}
G. Passarino and M. Veltman, ``One-loop corrections for $e^+e^-$
annihilation into $\mu^+\mu^-$ in the Weinberg model,'' Nucl. Phys. B
\textbf{160}, 151--207 (1979).

\bibitem{PascualTarrach1984}
P. Pascual and R. Tarrach, \emph{QCD: Renormalization for the Practitioner},
Lecture Notes in Physics \textbf{194} (Springer, Berlin, 1984).

\bibitem{PeskinSchroeder1995}
M. E. Peskin and D. V. Schroeder, \emph{An Introduction to Quantum Field
Theory} (Addison-Wesley, Reading, MA, 1995), Chap.~16.

\bibitem{Grozin2005}
A. G. Grozin, ``Lectures on QED and QCD,'' arXiv:hep-ph/0508242 (2005).

\bibitem{GrossWilczek1973}
D. J. Gross and F. Wilczek, ``Ultraviolet Behavior of Non-Abelian Gauge
Theories,'' Phys. Rev. Lett. \textbf{30}, 1343--1346 (1973).

\bibitem{Politzer1973}
H. D. Politzer, ``Reliable Perturbative Results for Strong Interactions?''
Phys. Rev. Lett. \textbf{30}, 1346--1349 (1973).


\bibitem{tHV}
G. 't Hooft and M. Veltman, ``Regularization and Renormalization of Gauge
Fields,'' Nucl. Phys. B \textbf{44}, 189 (1972).

\bibitem{BBDM}
W. A. Bardeen, A. J. Buras, D. W. Duke, and T. Muta, ``Deep inelastic
scattering beyond the leading order in asymptotically free gauge theories,''
Phys. Rev. D \textbf{18}, 3998--4017 (1978).

\bibitem{tHooft1973}
G. 't Hooft, ``Dimensional regularization and the renormalization group,''
Nucl. Phys. B \textbf{61}, 455--468 (1973).

\bibitem{Jackiw}
R. Jackiw, ``Functional evaluation of the effective potential,''
Phys. Rev. D \textbf{9}, 1686 (1974).

\bibitem{PaladiniSexton}
B. Paladini and J. C. Sexton, ``Asymptotic expansion of the lattice scalar
propagator in coordinate space,'' Phys. Lett. B \textbf{448}, 76--84 (1999),
arXiv:hep-lat/9805021.

\bibitem{CW}
S. Coleman and E. Weinberg, ``Radiative Corrections as the Origin of
Spontaneous Symmetry Breaking,'' Phys. Rev. D \textbf{7}, 1888 (1973).

\bibitem{Sobol}
I. M. Sobol', ``On the Distribution of Points in a Cube and the Approximate
Evaluation of Integrals,'' USSR Comput. Math. Math. Phys. \textbf{7}(4), 86
(1967).

\bibitem{Owen}
A. B. Owen, ``Scrambled Net Variance for Integrals of Smooth Functions,''
SIAM J. Numer. Anal. \textbf{34}(5), 1884 (1997).

\bibitem{SciPy}
SciPy Developers, \texttt{scipy.stats.qmc.Sobol}, SciPy Reference Guide
(accessed October 19, 2026).

\bibitem{GuadaNemevsek2020}
V. Guada and M. Nemev\v{s}ek, ``Exact One-Loop False Vacuum Decay Rate,''
Phys. Rev. D \textbf{102}, 125017 (2020).

\bibitem{AiCruzGarbrechtTamarit2020}
W.-Y. Ai, J. S. Cruz, B. Garbrecht, and C. Tamarit, ``Gradient Effects on
False Vacuum Decay in Gauge Theory,'' Phys. Rev. D \textbf{102}, 085001
(2020).

\bibitem{BaratellaGauge2025}
P. Baratella, M. Nemev\v{s}ek, Y. Shoji, K. Trailovi\'c, and L. Ubaldi,
``Revising the Full One-Loop Gauge Prefactor in Electroweak Vacuum
Stability,'' Phys. Rev. Lett. \textbf{134}, 011601 (2025).

\bibitem{Degrassi2012}
G. Degrassi, S. Di Vita, J. Elias-Mir\'o, J. R. Espinosa, G. F. Giudice,
G. Isidori, and A. Strumia, ``Higgs mass and vacuum stability in the Standard
Model at NNLO,'' JHEP \textbf{08}, 098 (2012).

\bibitem{Buttazzo2013}
D. Buttazzo, G. Degrassi, P. P. Giardino, G. F. Giudice, F. Sala, A. Salvio,
and A. Strumia, ``Investigating the near-criticality of the Higgs boson,''
JHEP \textbf{12}, 089 (2013).

\bibitem{Chigusa2018}
S. Chigusa, T. Moroi, and Y. Shoji, ``Decay Rate of Electroweak Vacuum in
the Standard Model and Beyond,'' Phys. Rev. D \textbf{97}, 116012 (2018).

\bibitem{DunneMin2005}
G. V. Dunne and H. Min, ``Beyond the thin-wall approximation: Precise
numerical computation of prefactors in false vacuum decay,'' Phys. Rev. D
\textbf{72}, 125004 (2005).

\bibitem{Shen2023}
J. Shen, P. Draper, and A. X. El-Khadra, ``Vacuum decay and Euclidean lattice
Monte Carlo,'' Phys. Rev. D \textbf{107}, 094506 (2023).

\end{thebibliography}
\end{document}